\pdfoutput=1
\documentclass[12pt,onecolumn,a4paper]{IEEEtran}
\usepackage{amsmath,amsfonts,amssymb}
\usepackage{graphicx}
\usepackage{cite}
\usepackage{subfigure}
\usepackage{multirow}
\usepackage{algorithm}
\usepackage{algorithmic}
\usepackage{mathtools}
\usepackage{hyperref}
\usepackage{verbatim}

\usepackage{color}

\IEEEoverridecommandlockouts

\newtheorem{theorem}{Theorem}
\newtheorem{definition}{Definition}
\newtheorem{lemma}{Lemma}
\newtheorem{corollary}{Corollary}

\newtheorem{example}{Example}
\newtheorem{remark}{Remark}

\newcommand{\set}[1]{\mathcal{#1}}
\newcommand{\defined}{\triangleq}
\newcommand{\Real}{{\mathbb{R}}}

\newcommand{\graph}{\set{G}}
\newcommand{\nodes}{\set{V}}
\newcommand{\edges}{\set{E}}
\newcommand{\N}{\set{N}}

\newcommand{\X}{\set{X}}
\newcommand{\R}{\set{R}}
\newcommand{\A}{\set{A}}
\newcommand{\B}{\set{B}}
\newcommand{\C}{\set{C}}

\newcommand{\M}{\set{M}}
\newcommand{\T}{\set{T}}
\newcommand{\W}{\set{W}}
\newcommand{\U}{\set{U}}
\newcommand{\Y}{\set{Y}}
\newcommand{\Z}{\set{Z}}

\newcommand{\h}{h}

\newcommand{\eht}[2]{#1 \to #2}
\newcommand{\tail}[1]{\mathrm{tail}(#1)}
\newcommand{\head}[1]{\mathrm{head}(#1)}
\newcommand{\sessions}{\set{S}}
\newcommand{\parent}{\pi}
\newcommand{\fd}{\textit{fd}}
\DeclareMathOperator{\An}{An}
\DeclareMathOperator{\Dom}{Dom}
\newcommand{\pde}{P$d$E}

\newcommand{\before}[1]{  {  \color{yellow} #1   }  }

\def\varphi{{\mu_{A}}}

\def\psi{{\mu_{C}}}
\def\larrow{{\: \longrightarrow_{A}\:}}
\def\rightarrowtail{{\: \longrightarrow_{B}\:}}
\def\twoheadrightarrow{{\: \longrightarrow_{C}\:}}

\title{Cut-Set Bounds on Network Information Flow}
\author{\IEEEauthorblockN{ Satyajit Thakor~\IEEEmembership{Member, IEEE},
    Alex Grant~\IEEEmembership{Senior Member,~IEEE} and~Terence
    Chan~\IEEEmembership{Member, IEEE}}\thanks{S. Thakor is with School of Computing and Electrical Engineering, Indian Institute of Technology Mandi. A. Grant is with Myriota Pty Ltd. T. Chan is with the Institute for
    Telecommunications Research, University of South Australia. The material in this paper was presented in part at the IEEE International Symposium on Information Theory, Seoul, South Korea, June/July 2009 \cite{ThaGraCha09}. This
    work was performed in part while S. Thakor was with the the
    Institute for Telecommunications Research, University of South
    Australia. A. Grant and T. Chan are supported in part by the
    Australian Research Council under Discovery Projects DP150103658.
    
}
    }

\begin{document}
\maketitle

\begin{abstract}
  Explicit characterization of the capacity region of communication
  networks is a long standing problem.  While it is known that network
  coding can outperform routing and replication, the set of feasible
  rates is not known in general.  Characterizing the network coding
  capacity region requires determination of the set of all entropic
  vectors. Furthermore, computing the explicitly known linear
  programming bound is infeasible in practice due to an exponential
  growth in complexity as a function of network size. This paper
  focuses on the fundamental problems of characterization and
  computation of outer bounds for multi-source multi-sink networks.
  Starting from the known local functional dependencies induced by the
  communications network, we introduce the notion of irreducible sets,
  which characterize implied functional dependencies. We provide
  recursions for computation of all maximal irreducible sets. These
  sets act as information-theoretic bottlenecks, and provide an easily
  computable outer bound for networks with correlated sources. We extend the notion of irreducible sets
  (and resulting outer bound) for networks with independent
  sources. We compare our bounds with existing bounds in the
  literature. We find that our new bounds are the best among the known
  graph theoretic bounds for networks with correlated sources and for
  networks with independent sources.
\end{abstract}

\section{Introduction}

The network coding approach introduced in~\cite{AhlCai00,LiYeu03}
generalizes routing by allowing intermediate nodes to forward coded
combinations of all received data packets. This yields many benefits
that are by now well
documented~\cite{YeuLi06,FraSol07,FraSol08,HoLun08}.  One fundamental
open problem is to characterize the capacity region and the classes of
codes that achieve capacity. The single session multicast problem is
well understood. Its capacity region is characterized by
max-flow/min-cut bounds and linear codes are optimal~\cite{LiYeu03}.

Significant complications arise in more general scenarios, involving
multiple sessions.  %Linear network codes are not
%sufficient~\cite{DFZ05,ChaGra08}.  Furthermore, 
A computable
characterization of the capacity region is still unknown. One approach
is to develop bounds as the intersection of a set of linear
constraints (specified by the network topology and sink demands) and
the set of entropy functions $\Gamma^*$ (inner bound), or its closure
$\bar{\Gamma^*}$ (outer bound)~\cite{Yeu02,SonYeu03,YeuLi06}.  An
exact expression for the capacity region does exist, again in terms of
$\Gamma^*$~\cite{YanYeu07}.  Unfortunately, this expression, or even
the bounds~\cite{Yeu02,SonYeu03,YeuLi06} cannot be computed in
practice, due to the lack of an explicit characterization of the set
of entropy functions for three or more random variables. 
The difficulties arising from the
structure of $\Gamma^*$ are not simply an artifact of the way the
capacity region and bounds are written. It has been shown that the
problem of determining the capacity region for multi-source network
coding is completely equivalent to characterization of
$\Gamma^*$~\cite{ChaGra08}.

One way to resolve this difficulty is via relaxation of the bound,
replacing the set of entropy functions with the set of polymatroids
$\Gamma$ (which has a finite, polyhedral characterization). This
results in a \emph{geometric bound} that is in principle computable
using linear programming~\cite{Yeu02}.  In practice however, the
number of variables and constraints in this linear program both
increase exponentially with the number of links in the network. This
prevents numerical computation for any meaningful case of interest. An
alternative approach is to seek \emph{graphical bounds} based on
functional dependence properties and cut sets in graphs derived from
the original communications network.

The more difficult problems of characterization and computation of
bounds for networks with correlated sources has received less
attention than networks with independent sources. For a few special
cases, necessary and sufficient conditions for reliable transmission
have been found. In particular, it was recently showed~\cite{Han11}
that the minimum cut is a necessary and sufficient condition for
reliable transmission of multiple correlated sources to \emph{all}
sinks. This result includes the necessary and sufficient
condition~\cite{Han80},~\cite{BarSer06} for networks in which every
source is demanded by single sink as a special case.  However, the
correlated source problem is an uncharted area in general. 
%To the best
%of our knowledge, there does not yet exist a nontrivial necessary
%condition for reliable transmission of correlated sources in general
%multi-sink networks.
% Is this really true?
%
A related important problem is that of separation of distributed
source coding and network coding.  It has been
shown~\cite{RamJaiChoEff06} that separation holds for two-source
two-sink networks. However it has also been shown by example that
separation fails for two-source three-sink and three-source two-sink
networks.

In this paper we develop new outer bounds for the capacity region of
general multicast networks with correlated sources. We further develop
the main concepts to also give tighter bounds for networks with
independent sources. The main idea of these bounds is to find subsets
of random variables in the network that act as \emph{information
  blockers}, or \emph{information-theoretic cut sets}. These are sets
of variables that determine all other variables in the network. We
develop the properties of these sets, which leads to recursive
algorithms for their enumeration. These algorithms can be thought of
as operating on a specially constructed functional dependence graph that
encodes the \emph{local} functional dependencies imposed by encoding
and decoding constraints.

\subsection{Organization and Main Results}
Section~\ref{sec:background} provides required background, including a
review of regions in the entropy space. These regions are used to
describe a family of geometric bounds on the
capacity region for network coding. We also describe
existing graphical bounds. %Appendix~\ref{sec:Rcs} extends the family of
%geometric bounds to handle correlated sources. 
Section \ref{sec:mainresults} presents main results of the paper.
In Section~\ref{sec:FDG}, we generalize the concept of a
functional dependence graph (FDG), Definition~\ref{def:FDG}, to handle
polymatroidal variables (a wider class of objects than random variables). 
This gives us a single framework that supports both
geometric and graphical bounds.  
%We give a basic result relating local
%and global dependence and establish in Theorem~\ref{thm:fd} that
%\fd-separation in a functional dependence graph implies (the
%appropriate generalization of) conditional independence for pseudo
%variables. 
Following on from this, we introduce the notion of
irreducible sets and maximal
irreducible sets for functional
dependence graphs, which are our key ingredients for characterization
and computation of capacity bounds.
Recursive algorithm finding all maximal irreducible sets for cyclic 
FDGss is developed using the structural properties of maximal
irreducible sets.
In Section \ref{sec:Multicasting}, we describe construction of a cyclic
FDG, called network FDG, from a given multi-source multi-sink network.
%with arbitrarily correlated sources. 
Maximal irreducible sets in a network
FDG are information bottlenecks, and provide  Theorem~\ref{thm:bound:b}\footnote{A simpler version of this bound was presented at IEEE International Symposium on Information Theory, Seoul, South Korea, June/July 2009 \cite{ThaGraCha09}} which outer bounds
the capacity region for networks with correlated sources.  It is established that
Theorem~\ref{thm:bound:b} is the best known graph theoretic bound
for multi-source multi-sink networks with correlated sources. 
In Section~\ref{sec:FDBdIS} we adapt our approach to take advantage of
the additional constraints introduced when sources are mutually
independent. This results in an improved bound, Theorem~\ref{thm:mainresult2}. In Appendix \ref{sec:Rcs} we give an algorithm to enumerate all maximal irreducible sets for acyclic FDGs. %,
%which, by Theorem~\ref{thm:RidsubsetRfd} is tighter (for independent
%sources) than Theorem~\ref{thm:mainresult1} (which holds for the more
%general correlated source case).
%
In Section~\ref{sec: comparison}, we compare our new bounds with
previously known results: cut-set bound~\cite{CovTho06}, network sharing bound~\cite{YanYan06}, the notion of information dominance \cite{HarKle06} and progressive \emph{d}-separating edge-set bound~\cite{KraSav06}.

\subsection{Notation}
Sets will be denoted with calligraphic typeface,
e.g. $\X$. Set complement is denoted by the superscript $\X^c$ (where
the universal set will be clear from context). Set subscripts identify
the set of objects indexed by the subscript: $X_\A=\{X_a,
a\in\A\}$. Collections of sets are denoted in bold, e.g.,
$\boldsymbol{\mathcal A}$. The power set $2^{\mathcal X}$ is the collection of all subsets of $\X$. Where no
confusion will arise, set union will be denoted by juxtaposition,
$\A\cup\B=\A\B$, and singletons will be written without braces.

%%%%%%%%%%%%%%%%%%%%%%%%%%
\section{Background}\label{sec:background}
 
\subsection{Poymatroids} \label{sec:pseudo-variables} 

We start with a brief review on classes of polymatroids which are used to derived a framework to characterize outer bounds on the network coding
capacity region. The framework will also enable us to understand the
connection between some geometric bounds and graphical bounds. As we shall see, some of these graphical bounds can be interpreted as relaxations of geometric bounds.

Let $\X$  be a set of $n$ variables and  $h$ be a  real-valued function 
$ h:  2^\X \mapsto\Real$
   such that 
$h(\emptyset)=0$. 
Each function $h$ can also be viewed as a
column vector in $\Real^{2^n}$ (or in $\Real^{2^n-1}$ knowing that $h(\emptyset)$ is always $0$) often called  the
entropy space~\cite{yeu97}.

\begin{definition}[Polymatroidal function or polymatroids]
A function
  $h:2^\X \mapsto\Real$ is \emph{polymatroidal} if it
  satisfies the following \emph{polymatroid axioms}
  \eqref{poly:zero}-\eqref{poly:submod} for all disjoint $\A,\B, \C \subseteq
  \X$.
  \begin{align}
    h(\emptyset) &= 0 \label{poly:zero} \\
 h(\B | \A) \triangleq h(\A \cup \B)-h(\A) & \ge 0         \label{poly:nondec}\\
 I_{h}(\A;\B | \C)\triangleq h(\A \cup \C)+h(\B\cup \C)-h(\A \cup \B \cup \C)-h(\C) & \ge 0.      \label{poly:submod}
  \end{align}
The set $\X$ is called the \emph{ground set} of the polymatroid $h$.  
\end{definition}
 
\begin{definition}[Entropic function]
A function $h$  is called \emph{entropic} if there exists a set of  $n$ discrete  random variables $(X_{v} : v \in \X)$ such that 
$$
h(\A)=H(X_{v} : v \in \A)
$$  
for all $\A \subseteq \X$. Here,  $H( \cdot )$ is the   Shannon entropy function.
\end{definition}

It is well known that all entropic functions are polymatroids. In the context of entropy functions, those polymatroid axioms are equivalent to  the \emph{basic}, or \emph{Shannon-type} inequalities~\cite{Yeu08}. For these reasons, an element in a ground set of a polymatroid may also be called ``variable''. Note that the chain rule for polymatroids also directly follows from the definition of $h(\cdot|\cdot)$. Functional dependency and independence in polymatroids can also be similarly defined as in random variables. Specifically,   with respect to a polymatroid $h$,  
\begin{enumerate}
\item
a subset of variables $\A$ is a function of another subset of variables $\B$ if 
\[
h(\A|\B) = 0,
\]
\item
a subset of variables $\A$ is conditionally independent of another subset of variables $\B$ given $\C$ if 
\[
I_{h}(\A ;\B | \C) = 0.
\]
\end{enumerate}

\begin{definition}[Almost entropic function]
  A function $h$ is \emph{almost entropic} if there exists a sequence
  of entropic functions $H^{(k)}$ such that $\lim_{k\to\infty} H^{(k)}
  = h$.
\end{definition}
 
%%%%%%%%%%%%%%%%%%%%%%%%%%%%%%%

Let $\Gamma$, $\Gamma^{*}$   and $\bar\Gamma^*$ be respectively the set of all polymatroidal, entropic and almost entropic functions. 
It is clear that 
\begin{align}
  \Gamma^* \subseteq \bar\Gamma^* \subseteq \Gamma. \label{eq:GammastarSubBarSubGamma}
\end{align}
In general, the region $\Gamma^*$ is not closed and hence  $\bar\Gamma^*$ strictly contains $\Gamma^* $. While $\bar\Gamma^*$ is convex~\cite{ZhaYeu97}, it is still extremely hard to characterize $\bar\Gamma^{*}$ (and hence also $\Gamma^*$). In fact, $\bar\Gamma^{*}$ is not even a polyhedron for $n > 3$ \cite{Mat07}. On the contrary, its outer bound  $\Gamma$ is a much simpler polyhedron in the non-negative orthant $\Real_{+}^{2^{n}-1}$ and in fact is the intersection of 
%a finite number of
%half-spaces, \eqref{poly:nondec}-\eqref{poly:submod}, which can be
%expressed minimally~\cite{yeu97} by 
\begin{equation}\label{eq: Gamma ineq}
  m = n + \binom{n}{2}2^{n-2}
\end{equation}
half spaces induced by the following elemental inequalities~\cite{yeu97}
\begin{align}
h(A | \X \setminus \{A\}) & \ge 0 \\
I_{h}(A ; B | {\C})& \ge 0 
\end{align}
where 
$A,B \in \X$ and $\C \subseteq \X \setminus \{A,B\}$.

%%%%%%%%%%%%%%%%%%%%%%%%%%%%%%%

\subsection{Network Coding}\label{sec:network coding}
Let the directed acyclic graph $\graph = (\nodes, \edges)$ serve as a
simplified model of a communication network with error-free
point-to-point communication links. We use  $\tail{e}$
and $\head{e}$ to respectively denote tail and the head of the directed edge $e$.   For nodes $u,v$ and edge $e$, we write $u\rightarrow e$  as a shorthand for $u=\tail{e}$ and
$e\rightarrow v$ for $v=\head{e}$. Also, for $e, f \in \mathcal E$, we write $e \rightarrow f$ if $\text{head}(e)=\text{tail}(f)$.  A \emph{path} in a directed graph
is a sequence of nodes $v_1,...,v_n$ such that there exists edges
$e_1,...,e_{n-1}$ with $\tail{e_i}=v_i$ and $\head{e_i}=v_{i+1}$. Such
a path is said to have length $n-1$.  Node $v_n$ is \emph{reachable}
from node $v_1$ if there exist a path from node $v_1$ to
$v_n$. Furthermore, node $v_n$ is \emph{connected} to $v_1$ if there
exist  nodes $v_1,...,v_n$ and edges
$e_1,...,e_{n-1}$ with $\head{e_i}=v_i$ and $\tail{e_i}=v_{i+1}$,
and/or $\tail{e_i}=v_{i}$ and $\head{e_i}=v_{i+1}$.  In other words, $v_n$ is connected  to $v_1$ if the two nodes are connected, by ignoring the direction of the edges.

%Each edge  $e\in\edges$ has a transmission capacity  $c_e>0$.  

Let $\sessions$ be an index set for multicast sessions  and   $\{Y_{s}: s \in
\sessions\}$ be the set of sources. The source $s$ is available at the set of  nodes $a(s)$  and is  demanded by multiple sink nodes  $b(s) \subseteq \nodes$. We call the tuple $(a,b)$ the connection requirement.  %To simplify our notations in the paper, we will assume that $a(s)$ is a singleton. And to study for the general case, we can construct a ``super source node'' (for each source) with directed edges (of unlimited capacities) to all the source nodes $a(s)$.

In this paper, we assume that the sources are  i.i.d. sequences
$$
\{(Y_{s}^{n}, s\in\sessions) , \: n = 1, 2, \ldots, \} 
$$
so that copies of $(Y_{s}^{n}, s\in\sessions)$ generated at different time $n$ will be  independent of each other. However, within the same time instance $n$, the sources $(Y_{s}^{n}, s\in\sessions)$ may be correlated among different sources.  In the special case when $(Y_{s}^{n}, s\in\sessions)$ is also mutually independent, we will say the sources are independent.  
Also, the distribution of $(Y_{s}^{n}, s\in\sessions)$ and hence entropies of any subset of sources are assumed to be known. For notation simplicity, we will use  $(Y_{s}, s\in\sessions)$ to denote a generic copy of the sources at any particular time instance.

For a network $\graph=(\nodes,\mathcal E)$ subject to a connection requirement
$a$ and $b$, a deterministic network code (of block length $n$) 
 is a collection of source and edge random variables $(Y_{s}^{[n]}, s\in\sessions, U_{e}^{(n)}, e\in\edges)$ where 
$U_e^{(n)}$ is  the message transmitted on the edge  $e\in\edges$  and $ Y_{s}^{[n]}$ is the block of source symbols
$
(Y_{s}^{1}, \ldots, Y_{s}^{n})
$.
Unlike $ Y_{s}^{[n]}$ which is a collection of $n$ i.i.d. random variable,  the superscript $(n)$ in $U_{e}^{(n)}$ is only used to indicate the block length of the code. It does not mean that $U_{e}^{(n)}$  is a collection of $n$ i.i.d. random variables.

Clearly, these random variables $(Y_{s}^{[n]}, s\in\sessions, U_{e}^{(n)}, e\in\edges)$ cannot be arbitrarily but must satisfy some constraint. In particular, it is required that  1)   an edge random variable must be a function of incident edge random variables and source random variables, and 2) for any $s\in \sessions$, a sink node $v \in b(s)$ must be able to reconstruct the demanded source. More precisely, we have the following definition.

\begin{definition}[Network code]\label{def:networkcode}
%With respect to a given network   $\graph=(\nodes,\mathcal E)$ and sources $(Y_{s}, s\in\sessions)$, a 
A network code $\phi_{\graph}^{(n)}=\{\phi_{e}^{(n)},\phi_{u,s}^{(n)}\}$ of block length $n$  is described by a set of local encoding
  functions $\phi_{e}^{(n)},e \in \mathcal E$ and decoding functions
  $\phi_{u,s}^{(n)}, u \in b(s), s \in \sessions$ 
  \begin{align*}%\label{eq:networkcode}
\phi_{e}^{(n)}&: \prod_{j \in \sessions : j
        \rightarrow e} \mathcal Y_{j}^{[n]} \times \prod_{f \in \mathcal E:
        \eht{f}{e}} \U_{f}^{(n)}   \longmapsto \U_{e}^{(n)},\\% , e \in \mathcal E,\\
    \phi_{u,s}^{(n)}&:\prod_{j \in \sessions : j
        \rightarrow u} \mathcal Y_{j}^{[n]} \times \prod_{f \in \mathcal E: \eht{f}{u}} \mathcal
      U_{f}^{(n)} \longmapsto \mathcal Y_{s}^{[n]}.%, u \in b(s), s \in \sessions.
  \end{align*}  
Here, the alphabets  of the block of source random variables $Y_{s}^{[n]}$ 
and
edge random variables $U_{e}^{(n)}$ are denoted by $\mathcal Y_{s}^{[n]}$ and $\U_{e}^{(n)}$  respectively. 
\end{definition}

\begin{remark}
With respect to a given network code, the joint distribution for the set of all source and edge random variables $(Y_{s}^{[n]}, U_{e}^{(n)}, s\in\sessions, e\in\edges)$ will become well-defined. Furthermore, for any $e\in\edges$, one can construct a global encoding function  such that 
\begin{equation}\label{eq:globalcode}
U_{e}^{(n)} = \tilde{\phi}_e^{(n)} (Y_{s}^{[n]}, s\in\sessions) 
\end{equation}
\end{remark}

\begin{definition}[{Achievability}]
\label{def:Achievable rate tuple1}
  An edge capacity tuple $\mathbf{c}=\left(c_{e}:e \in \edges\right) \in
  \Real_{+}^{|\edges|}$ is called \emph{achievable}  if there exists a sequence of network codes 
  $$
  \phi_{\graph}^{(n)} = \{\phi_{e}^{{(n)}}, \phi_{u,s}^{{(n)}}, e\in\edges,s\in\sessions, u\in b(s)\}
  $$ 
  (and also the corresponding induced  source and edges random variables  
  $
  (Y_{s}^{[n]}, U^{(n)}_{e}, s\in \sessions, e\in\edges)
  $)
such that 
  \begin{align*}
\limsup_{n\to\infty}      {\log_{{2}} \left|\U^{(n)}_{e}\right|}/{n} & \leq  c_{e},\\  %\label{eq:sequenceofNC1}%\\
    %
%  \end{align*}
 % \begin{align*}
  \limsup_{n\to\infty}  \mathrm{Pr}\left\{
  \phi_{u,s}\left(Y_{j}^{[n]},U^{(n)}_{f}: f
    \to u, j \to u\right) \neq Y^{[n]}_s\right
    \} &= 0%\label{eq:sequenceofNC2}%, \forall u \in b(s).
  \end{align*}
for all $e \in \mathcal E$ and $u \in b(s)$.
\end{definition}

\begin{remark}
When sources are correlated, it is natural to assume a fixed joint distribution of the sources. In that case, the network coding capacity region $\R$ is the set of all achievable edge capacity tuples that support the transmission of the sources. 
When sources are independent, only the entropies  but not the joint distribution matter (as one can always compress the sources independently before transmission).  Therefore, as in some existing literature, one may instead focus on finding the set of source rates or entropies that a network can transport, subject to a fixed edge capacity tuple.
\end{remark}

%%%%%%%%%%%%%%%%%%%%%%%%%%%%%%%%
%\subsection{Bounds for networks}\label{sec:geo_bounds}
\subsection{Network Coding Bounds}\label{sec:geo_bounds}

%Let $\R $ be the set of all achievable edge capacity tuples. 
Definition~\ref{def:pseudo-variable bounds} below provides a
standard framework to formulate ``geometric'' bounds on the set of achievable edge capacity  tuples (denoted by $\R$).
%
% in terms of variables~\cite{Yeu08}.
\begin{definition}\label{def:pseudo-variable bounds}
Consider any network coding problem (with an underlying network 
 $\graph=(\nodes,\edges)$  and connection requirement  $(a,b)$). 
  For any non-empty subset  $\Delta $  of   polymatroids   on the ground set $\X=(Y_{s}, U_{e}, s\in\sessions, e\in \edges)$,  let $\R(\Delta)$ be the set
  of tuples
  $(c_{e},e\in\edges)\in \Real^{|\edges|}$ for
  which there exists $h\in\Delta$ satisfying
  \begin{align}
    h\left(Y_s:s \in \A \right)-H\left(Y_s:s \in \A \right) &=  0, \: \A \subseteq \sessions \label{eq:source input} \\
    h\left(U_e \mid  Y_j  , {\eht{j}{e}} , \:  U_f, \eht{f}{e}  \right) & = 0, \:   e\in\edges \label{eq:Networkcoding}\\
    h\left(Y_s \mid  Y_{j}, j \to u , \: U_f, \eht{f}{u} \right) & = 0, \: u\in b(s), s\in\sessions \label{eq:networkdecoding}\\
    h\left(U_e\right) & \le c_e, \: e\in\edges \label{eq:CapacityConstraint}  %
%          h\left(Y_s \right) & \ge r_{s}, s\in\sessions 
%
%    h\left(Y_s\right) & \ge R_s, s\in\sessions \label{eq:LastConstraint}
  \end{align}
\end{definition}

{
\begin{remark}
Note that, $Y_{s}$ in \eqref{eq:source input} can be viewed as a generic source random variable and also as an element in the ground set $\X$. 
\end{remark}
}

Here we can identify constraints due to source correlation~\eqref{eq:source input}, network coding~\eqref{eq:Networkcoding},
decoding~\eqref{eq:networkdecoding}, edge
capacity~\eqref{eq:CapacityConstraint}. 
% and source rate~\eqref{eq:LastConstraint} constraints.  
Each of these constraints
defines a region  of polymatroids 
\begin{align}
  \C_1 &\triangleq\{\h: \h\textrm{ satisfies }\eqref{eq:source input}\} \\
  \C_2 &\triangleq\{\h: \h\textrm{ satisfies }\eqref{eq:Networkcoding}\} \\
  \C_3 &\triangleq\{\h: \h\textrm{ satisfies }\eqref{eq:networkdecoding}\} \\
  \C_4 &\triangleq\{\h: \h\textrm{ satisfies }\eqref{eq:CapacityConstraint}\}.
\end{align}

When sources are independent, i.e., $ h\left(Y_s:s \in \sessions \right)= \sum_{s \in \sessions}h\left(Y_s\right)$,  $\R(\Gamma^*)$ and $\R(\bar\Gamma^*)$ are respectively  inner and outer bounds for $\R$~\cite[Chapter 15]{Yeu02}. In Yan et al.~\cite{YanYeu07},  an
exact characterization of   $\R$ for multi-source
multi-sink network coding was also obtained. 

{
When sources are correlated, using arguments similar to those used in the proof of \cite[Theorem 15.9]{Yeu02}, one can prove that  $\R(\bar\Gamma^*)$ is still an outer bound for $\R$. Note that in the bound  $\R(\bar\Gamma^*)$, only the joint entropies of the sources but not their joint probability distribution are used to derive the bound. Therefore, one can tighten the bound by 
incorporating additional information about  the joint distribution in characterizing bounds (see \cite{GohYanJagg11}, \cite{ThaChaGra11} and \cite{ThaChaGra13}).
 
%
%\stc{
%Such outer bounds may be viewed as bounds on the necessary and sufficient matching conditions between correlated sources and link capacities. The resulting bounds are not tight and can be tightened by
%incorporating the knowledge of joint distribution in characterizing bounds
%\cite{GohYanJagg11}, \cite{ThaChaGra11} (see also \cite{ThaChaGra13}).}
%%

Since entropy functions and almost entropic functions are
polymatroidal~\eqref{eq:GammastarSubBarSubGamma} and the regions
$\bar\Gamma^*$, $\Gamma$, $\C_1$, $\C_2$, $\C_3$, $\C_4$ are closed
and convex, it follows that $\R(\Gamma)$ is an outer bound for the set
of achievable rates. The
relation of these capacity bounds is summarized below.
\begin{equation}\label{eq:bounds}
\R \subseteq \R(\bar\Gamma^*) \subseteq \R(\Gamma)
\end{equation}

Weighted sum-rate bounds induced by $\R(\Gamma)$ can in principle be computed using linear
programming. One practical difficulty with numerical computation of such 
bounds is that the number of variables and the number of
constraints due to $\Gamma$ both increase exponentially
with $|\sessions|+|\edges|$ (refer to~\eqref{eq: Gamma ineq}).   Attempts to simplify these
bounds using direct application of Fourier-Motzkin~\cite{Sch98} may prove
fruitless. In \cite{ThaGraCha11}, the authors have proposed a graph based approach to simplify the bound by exploiting the abundant set of functional dependencies in a network coding problem.
}

\begin{comment}
\before{
NOT SURE if we want to keep this or not. Or where should move it to.

For convenience we describe existing graph theoretic bounds
(i.e., bounds that rely on a graph representation of the network) for
network information flow in Section \ref{sec: comparison} where we also compare our main results with these bounds. We consider the cut-set
bound~\cite{CovTho06}, network sharing bound~\cite{YanYan06} and
progressive $d$-separating edge-set bound~\cite{KraSav06}. We
also review the idea of information dominance described
in~\cite{HarKle06} (from which we later obtain an ``information
dominance bound'', Theorem~\ref{thm:IDbound}). 
}
\end{comment}

In addition to above bounds, there are also many ``graphical'' bounds (i.e., bounds that rely on a graph representation of the network coding system)  in existing literatures. We will review and compare these bounds, such as   
cut-set
bound~\cite{CovTho06}, network sharing bound~\cite{YanYan06} and
progressive $d$-separating edge-set bound~\cite{KraSav06}
in Section \ref{sec: comparison}.

%%%%%%%%%%%%%%%%%%%%%%%%%%%%%%%%%%%
\section{Main Results}\label{sec:mainresults}
%\subsection{Graphical Bounds for Networks with Correlated Sources}
%\label{sec:FDBound}
The main results of this paper are graphical bounds for networks with correlated or independent sources.
In Section~\ref{sec:FDG}
we will define a functional dependence graph, which represents a set of
local functional dependencies between polymatroidal variables. Our definition
extends~\cite{Kra98} to accommodate cycles containing source nodes,
and polymatroidal variables in place of random variables. This section also
provides the main technical ingredients for our new bounds. In
particular, we describe a test for functional dependence, and give a
basic result relating local and global dependence.
Section~\ref{sec:Multicasting} describes our new bound for general multicast networks with
correlated sources, based on the implications of local functional
dependence. Section~\ref{sec:FDBdIS} considers source independence implications to further strengthen the proposed bound.

{
 The main ingredient of most graph based outer bounds is the following theorem:
\begin{theorem}[Bottleneck Bound]\label{thm:1}
  \label{thm:mainresult1}
  Let $\B=\{U_{\A},Y_{\W^c}\}$ be a set such that 
  \begin{align}\label{eq:thm1}
  h(\B) = h(Y_{s}, s\in\sessions ) 
  \end{align}
for any polymatroid $h \in  \C_{1}\cap \C_2\cap\C_3 \cap\C_4$.  
%for any  $h \in \R(\Gamma)$. 
Then, 
  \begin{equation}\label{eq:mainresult1}
      \sum_{e \in \A} c_e \geq  H\left(Y_{\W}\mid Y_{\W^c}\right).
  \end{equation}
\end{theorem}

\begin{IEEEproof}
Notice that  
\begin{align*}
     H\left(Y_{\W}\mid Y_{\W^c}\right) 
    & =  h\left(Y_{\W}\mid Y_{\W^c}\right)  \\
    &= h\left(Y_{\sessions}\right) - h\left(Y_{\W^c}\right) \\
    &= h(\B) - h\left(Y_{\W^c}\right)\\
    &= h\left(U_{\A}\mid Y_{\W^c} \right) \\
    & \leq \sum_{e \in \A} h(U_e)\\
    & {\leq} \sum_{e \in \A} c_e
  \end{align*}
and the theorem is proved.  
\end{IEEEproof} 

As a consequence, one may identify various subsets $\B$ satisfying \eqref{eq:thm1} and use them to derive bounds for the network coding rate region. The question however is how to find such bottleneck subsets. Finding all bottlenecks can be a very challenging and computing intensive task. 
 In the remaining of the section, we will derive various graph based technique to find such bottlenecks.
}
%%%%%%%%%%%%%
\subsection{Functional Dependence Graphs}\label{sec:FDG}

{
\begin{definition}[Functional Dependence Graph]\label{def:FDG}
  Let $\Delta$ be a set of polymatroids on a ground set $\X=\{X_1,\dots,X_N\}$. A directed graph
  $\graph^{*}=(\X,\edges^{*})$ is called a
  \emph{functional dependence graph} for $\Delta$ if and only if for all
  $i=1,2,\dots,N$
  \begin{equation}\label{eq:localdependence}
    h\left(X_i \mid   X_j : (j,i)\in\edges^{*}  \right) = 0, \forall h \in\Delta
  \end{equation}
Alternatively, a function $h$ is said to satisfy the FDG  $\graph^{*}=(\X,\edges^{*})$ if it satisfies \eqref{eq:localdependence}.
 An FDG is  called \emph{cyclic} if every node is a member of a directed
cycle. 
 
\end{definition}
}
 
Definition~\ref{def:FDG} is more general than the FDG of~\cite[Chapter 2]{Kra98}:  Firstly, in our definition there is no distinction between source and non-source
random variables. The graph simply characterizes functional dependence
between variables. In fact, our definition admits cyclic directed
graphs with cycles containing source nodes, and there may be no nodes
with in-degree zero (which are source nodes in~\cite{Kra98}). We also
do not require independence between sources (when they exist), which
is implied by the acyclic constraint in~\cite{Kra98}. Our definition
admits functions $h$ with \emph{additional} functional
dependence relationships that are not represented by the graph. It
only specifies a certain set of conditional functions which
must be zero. Our definition holds for a wider class of
objects (variables in polymatroids) rather than only random variables.
Clearly an FDG in the sense of~\cite{Kra98}
satisfies the conditions of Definition \ref{def:FDG}, but the converse
is not true. 
For clarity, a functional dependence graph (FDG) is  defined according to our 
 Definition~\ref{def:FDG}.

Definition \ref{def:FDG} specifies an FDG in terms of local
dependence structure. Given such local dependence constraints, it
is of great interest to determine all implied functional
dependence relations. In other words, given an FDG, we wish to
find all sets $\A$ and $\B$ such that $h(\B|\A )=0$ for all $h$ satisfying the FDG.

 {
\begin{definition}[$\A$ determines $\B$]\label{def:fd}
Consider a directed graph $\graph^{*}=(\X,\edges^{*})$. 
For any sets $\A,\B \subseteq \X$, we say that $\A$ determines $\B$ (with respect to Procedure A) if there are no elements of $\B$ remaining after
  the following procedure: 
  
  {\bf Procedure A: }
  \begin{quote}
  { Remove all  the edges outgoing from the nodes in $\A$ and subsequently remove all   nodes and edge with no incoming edges and nodes respectively}.
  \end{quote}
  
We will use  $\A\larrow\B$ to denote that   $\A$ determines $\B$.
\end{definition}

\begin{definition}[Blanket]\label{def:phi}
  For a given set $\A$, let $\varphi(\A)\subseteq\X$ be the set of
  nodes deleted by the procedure of Definition~\ref{def:fd} together with the nodes in $\A$. We will call $\varphi(\A)$ the \emph{blanket} of $\A$ (with respect to Procedure A).
\end{definition}
}

Clearly $\varphi(\A)$ is the largest set of nodes with
$\A\larrow\varphi(\A)$. To this end, define for $X_{i} \in \X$
\begin{equation}\label{eq:parent}
  \parent(X_{i}) = \{X_{j}\in\nodes: (X_{j} ,X_{i})\in\edges^{*}\}
\end{equation}
to be the set of parents of node $X_{i}$. Where it does not cause
confusion, we will abuse notation and identify variables
and nodes in the FDG, e.g. \eqref{eq:localdependence} will be
written
  $h\left(X_{i} \mid \parent(X_{i})\right) = 0$ or simply 
    $h\left(i \mid \parent(i)\right) = 0$.
 
%\stc{
%\begin{remark}
%Note that, 
%\end{remark}
%}

\begin{lemma}[Grandparent lemma]\label{lem:grandparent}
  Let $\graph^{*}=(\X,\edges^{*})$ be an FDG for a
  polymatroid $h$. For any
  $j\in\nodes$ with $i\in\parent(j)\neq\emptyset$
\begin{equation}
  h\left(j\mid \parent(i), \parent(j) \setminus i \right) = 0.
\end{equation}
\end{lemma}
\begin{IEEEproof}
  By hypothesis, $h(j\mid \parent(j))=0$ for any
  $j\in\nodes$. Furthermore, note that for any $h\in\Gamma$,
  conditioning cannot increase the function $h$\footnote{This is a
    direct consequence of submodularity  \eqref{poly:submod}.} and
  hence $h(j\mid\parent(j),\A)=0$ for any $\A\subseteq\X$. Now
  using this property, and the chain rule for polymatroids,
  \begin{align*}
    0 &= h(j\mid\parent(j)) \\
    &= h(j \mid \parent(j), \parent(i)) \\
    &= h(j, \parent(j), \parent(i)) - h(\parent(j), \parent(i)) \\
    &= h(j, \parent(j)\setminus i, \parent(i))
%+ \underbrace{g(i \mid j, \parent(j)\setminus i, \parent(i))}_0
- h(\parent(j), \parent(i)) \\
  &= h(j, \parent(j)\setminus i, \parent(i)) -
  h(\parent(j)\setminus i, \parent(i))
%-  \underbrace{g(i\mid\parent(j)\setminus i, \parent(i))}_0
\\
  &= h(j\mid \parent(i), \parent(j) \setminus i ).
  \end{align*}
\end{IEEEproof}
We emphasize that in the proof of Lemma~\ref{lem:grandparent}, we have
only used the submodular property of polymatroids, together with the
hypothesized local dependence structure specified by the FDG.
%By invoking the lemma recursively, we have the following  lemma. 
%
%Clearly the lemma is recursive in nature. For example, it is valid for
%$h(j\mid \parent(j) \setminus i, \parent(i)\setminus k, \parent(k))=
%0$ and so on. The implication of the lemma is that a variable
%$X_j$ in an FDG is a function of $X_\A$ for any $\A\subseteq\nodes$ with
%$\A\larrow j$.

\begin{lemma}\label{thm:fd}
Let $\graph^{*}=(\X , \edges ^{*})$ be an FDG for a polymatroid $h$. Then for disjoint
subsets $\A,\B\subseteq\X$,
\begin{equation}
\A \larrow \B \implies h(\B \mid \A) = 0.
\end{equation}
\end{lemma}
\begin{IEEEproof}
  Let $\A\larrow\B$. Then, by Definition
  \ref{def:fd} there must exist directed paths from some nodes in $\A$
  to each node in $\B$, and there must not exist a directed path
  to any node in $\B$ which does not also intersect $\A$. {In other words, apart from the paths from nodes in $\A$ and their sub-paths, any other path leading to $\B$ must have an element of $\A$ as its member.} Recursively
  invoking Lemma~\ref{lem:grandparent}, the lemma is proved.
\end{IEEEproof}

\begin{definition}[Irreducible set]\label{def:IrredSet1}
  A set of nodes $\B$  is 
  \emph{irreducible} (with respect to Procedure A) if there is no $\A\subseteq\B$ such that    $\A\larrow\B$. Furthermore, an irreducible set $\A$ is \emph{maximal}  if $ \varphi(\A)=\X$. 
\end{definition}

{
\begin{remark}
In this paper, we are mainly interested in cyclic FDGs to characterize cut-set bounds on network capacity. However, for other applications, acyclic FDGs may also be of interest. In Appendix \ref{sec:Rcs} we define maximal irreducible sets for acyclic network and give an algorithm to compute them.
%\end{remark}
%
%\begin{remark}
For cyclic graphs, every subset of a maximal irreducible set is
irreducible.  In contrast to acyclic graphs the converse is not true, that is, there can be irreducible sets that are not maximal and
are not subsets of any maximal irreducible set.
\end{remark}

\begin{corollary}\label{lem:equal}
If $\A$ and $\B$ are both maximal irreducible sets, then 
$h(\A)=h(\B)=h(\X)$ for any polymatorids satisfying the FDG $(\X,\edges^{*})$.
\end{corollary}
\begin{IEEEproof}
By Definition~\ref{def:IrredSet1}, $\varphi(\A)=\varphi(\B)=
  \X$. Invoking Lemma~\ref{thm:fd},  $h(\A)=h(\B)=h(\X)$.
\end{IEEEproof}

%\tc{Terence: Can you rewrite this part related to the two algorithms? Make sure I did it correctly.}

%\tc{
As we shall see, the corollary, together with Theorem \ref{thm:1},  can be used to derive  capacity bounds for
network coding.  Therefore, we are interested in finding every maximal
irreducible set. This may be accomplished via \textbf{AllMaxSetsC}($\graph^*_{\N},\{\}$) in Algorithm~\ref{alg:AMS}, which recursively  finds all maximal irreducible sets.  {In the algorithm, the graph $\graph^*_{\N}=(\X_{\N},\edges^*_{\N})$, where ${\N}=\{1,\ldots, {|\X|}\}$, is isomorphic to $\graph^*=(\X,\edges^*)$ via some bijection $\sigma: \X \longmapsto \X_{\N}$ and hence $(u,v) \in \edges^*$ iff $(\sigma(u),\sigma(v)) \in \edges^*_{\N}$. % } 
%that \emph{do not} contain any node in $\A$ given that the set $\A^c$ contains a maximal irreducible set. %Similar to Algorithm \ref{alg:augment}, 
 %For example, for $\graph$ with $|\nodes|=n$ and $|\edges|=0$, the function \textbf{AllMaxSetsC}($\cdot,\cdot$) is called only once.
{For set $\A \subseteq \N$ we define $\A^{'} =\{i \in \N : i > j, \forall j \in \A \}$.

\begin{algorithm}
  \caption{\textbf{AllMaxSetsC}($\graph^*_{\N},\A$)}\label{alg:AMS}
  \begin{algorithmic}[1]
    \REQUIRE $\graph^*_{\N}=(\N,\edges^*_{\N}), \A\subseteq\N$
%     \STATE $\A=\N$  
     \IF{$i\not\in\varphi\left(\A^c\setminus\{i\}\right), \forall i\in \A^c$} \label{stp:A2-1}
%     \IF{$\nodes \subseteq \varphi\left(\A^c\right)$}
    \STATE Output $\A^c$
%    \ELSE
%    \STATE{Return ``$\A^c$ does not contain maximal irreducible sets''}
%    \ENDIF
    \ELSE
                \FORALL{$i\in\A^{'}$}
                \IF{$i\in\varphi\left(\A^{'}\setminus\{i\}\right)$}
                \STATE Output $\textbf{AllMaxSetsC}(\graph^*_{\N},\A\cup\{i\})$ \label{stp:A2-6}
                    \ENDIF
             \ENDFOR
     \ENDIF
  \end{algorithmic}
\end{algorithm}

The actual number of operations (or the time complexity) to execute the function call depends on the topology of the FDG. The recursion tree is described in Figure \ref{fig:RecursionTree}. We make the following observations: (1) the leaf nodes of the recursion tree (such nodes are represented within circles) are subsets containing $|\mathcal X|$ and/or complement of maximal irreducible sets (denote by $\B$ a maximal irreducible set and by $\boldsymbol{\mathcal M} $ the set of all such sets), (2) any leaf node which is not a complement of any $\B \in \boldsymbol{\mathcal M}$ is a subset of some $\B^c, \B \in \boldsymbol{\mathcal M}$ and (3) each node of the recursion tree represents a unique set. Hence the number of nodes are upper bounded by the cardinality of the set $\cup_{\B \in \boldsymbol{\mathcal M}} 2^{\B^c}$. Using the union bound, the total number of calls of the function \textbf{AllMaxSetsC}($\graph^*_{\N},\{\}$) can be upper bounded by
\begin{align*}
\sum_{\B \in \boldsymbol{\mathcal M}} 2^{|\X|-|\B|}.
\end{align*}

\begin{remark}
Due to the recursive nature, the algorithm is easy to implement. The number of recursive calls can be further reduced, for example, by providing all cut-sets separating subsets of sources and corresponding sinks and using complement of the cut-sets as input to Algorithm \ref{alg:AMS} while replacing $\A^{'}$ by $\A^c$ (this is important for input other than $\{\}$). We will see in Section \ref{sec: comparison} that the maximal irreducible sets are subsets of such cut-sets. 
\end{remark}

\begin{figure}[htbp]
  \begin{center}
    {\includegraphics[scale=.63]{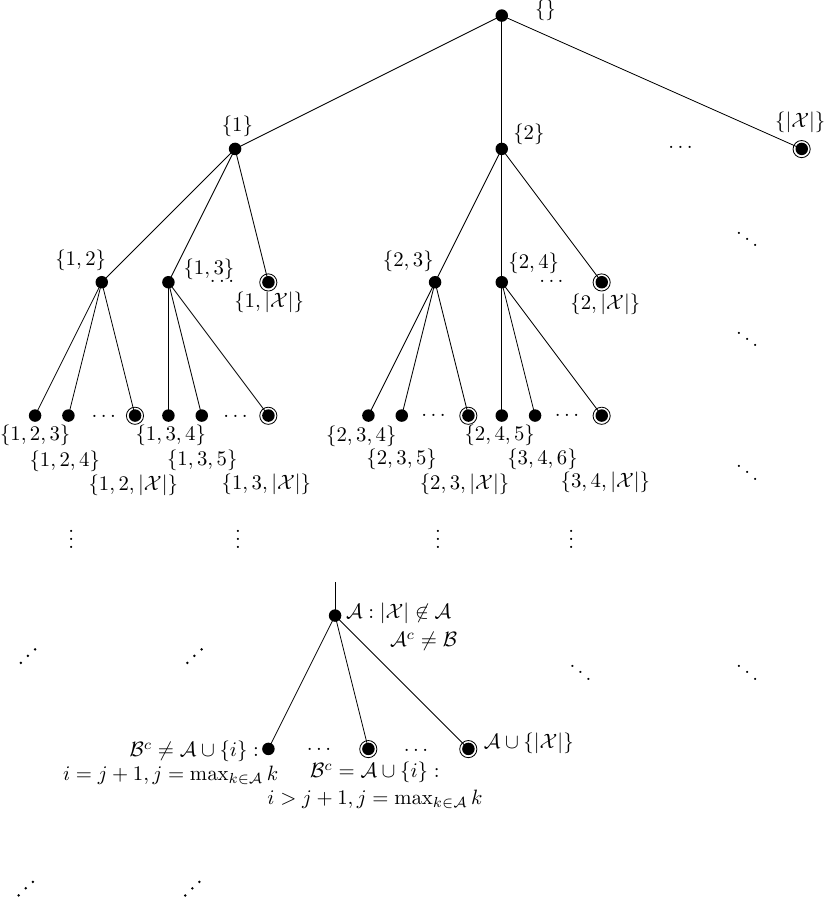}}
    \caption{Recursion tree, $\B$ is any maximal irreducible set.}\label{fig:RecursionTree}
  \end{center}
\end{figure}
}
}
}

\subsection{A Bound for Network with Correlated Sources}\label{sec:Multicasting}

So far,  we have defined functional dependence graphs, developed some of their properties, and given algorithm for finding all maximal
irreducible sets. In order to apply these results to find bounds on
network coding capacity, we  need to construct FDGs from multi-source
communications networks with multicast constraints.
 
{
\begin{definition}[Network FDG]\label{def:FDG-B}
For a given network coding problem (defined by the network topology $\graph$ and connection requirement $(a,b)$),  its induced \emph{network FDG} is a directed graph $\graph^*=(\X , \edges^{*})$ defined as follows 
  
\begin{itemize}
\item

The set of nodes $\X$ is equal to  
$$
\{ U_{e}, e\in\edges \} \cup
\{ Y_{s}, s\in\sessions\} \cup  \{ \hat Y_{s}^{i}, s\in\sessions, i \in b(s)\},
$$  

\item
 $(A,B)$ is a directed edge in $\edges^{*}$ if it satisfies one of the following conditions
\begin{enumerate}
\item $A = U_{e}$, $B = U_{f } $ and $e \to f$;
\item $A = Y_{s}$, $B = U_{f } $ and $s \to f$;
\item $A = U_{e}$, $B = \hat Y_{s }^{i} $, $i\in b(s)$ and $e \to i$;
\item $A = Y_{\ell}$, $B = \hat Y_{s }^{i} $,  $i\in b(s)$ and $ i \in a(\ell)$;
\item $A = \hat Y_{s }^{i} $,  $B = Y_{s}$, and $i\in b(s)$.  
\end{enumerate}

\end{itemize} 
 
\end{definition}

 \begin{remark}
In the above definition, the physical meaning of  $\hat Y^{i}_s$ is the decoded estimates of $Y_s$ at
the sink node $i \in b(s)$.  
 Note that the decoding
constraints~\eqref{eq:networkdecoding} require that
  $Y_s=\hat Y^{i}_s, i \in b(s)$ for all $s \in
\sessions$.
 \end{remark}  
}

\begin{figure}[htbp]
  \begin{center}
    \subfigure[\label{butterfly}]
    {\includegraphics[scale=.9]{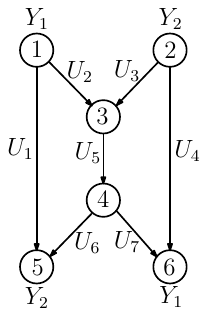}}
    \hspace{10mm} \subfigure[\label{mul2}]
    {\includegraphics[scale=.9]{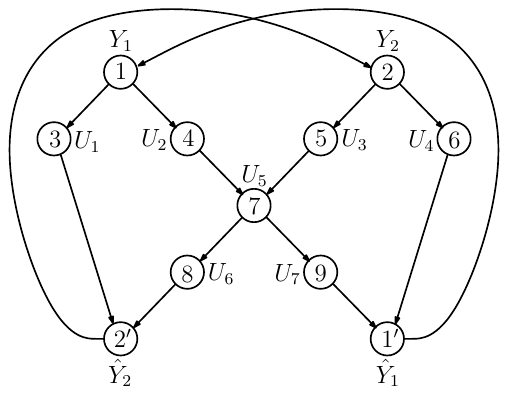}}
    \caption{The butterfly network (a) and its network FDG (b).}
  \end{center}
\end{figure}

\begin{example}[Network FDG of the butterfly network]
  Figure \ref{butterfly} shows the well-known butterfly network and
  Figure~\ref{mul2} shows its network FDG. Nodes are labeled with node numbers and variables. 
  Edges in the network FDG represent dependencies due to encoding
  and decoding requirements. 

\end{example}

 In network FDGs, there are nodes for auxiliary variables which represent decoding estimate and are the same as the source variables demanded at the sink. Accordingly, the following procedure finds functional dependency in network FDG taking multicasting into consideration. 

%We now extend our technical machinery in order to obtain 
%functional dependence bound from network FDG.  

{
\begin{definition}[Procedure B]
  \label{def:fd_HFDG}
  Consider a network FDG as defined in Definition \ref{def:FDG-B}.  
%  For any $\B \subseteq \A \subseteq \nodes$,  $\A$ determines $\B$. 
  For any sets $\A,\B\subseteq\X$, we say $\A$ determines $\B$
(with respect to  Procedure B) if there are no elements
  of $\B$ remaining after the following procedure:
  
{\bf Procedure B: }

\begin{quote}
  \begin{enumerate}
  \item Remove all edges outgoing from nodes in $\A$ and subsequently
    remove all nodes and edges with no incoming edges and nodes
    respectively.
  \item If any $\hat Y^i_s$ is removed, (a) remove all $\hat Y^i_s$ for $i \in \{1,\ldots,|b(s)|\}$ and (b) subsequently remove all edges and
    nodes with no incoming edges and nodes, go to Step 2. Else
    terminate.
\end{enumerate}
\end{quote}

We will use $\A\rightarrowtail\B$ to denote that  $\A$ determines $\B$
 with respect to  Procedure B. 
\end{definition}

As before, concepts such as  blanket and irreducibility  can be similarly defined with respect to Procedure B. Specifically, for a given set $\A$, 
its blanket (with respect to Procedure B) is  denoted by  
 $\mu_B(\A)$ and is defined as  the largest set of nodes with
$\A\rightarrowtail \mu_B(\A)$.
  A set of nodes $\B$  is called
  \emph{irreducible} (with respect to Procedure B) if there is no $\A\subseteq\B$ such that
  $\A\rightarrowtail\B$. An irreducible set $\A$ is \emph{maximal}  if
  $\X\setminus\mu_B(\A)=\emptyset$.
 In addition, if $\A$ and $\B$ are maximal irreducible sets, then 
\[
h(\A) = h(\B)
\]
for all polymatroid $h$ satisfying the network FDG.

Furthermore, the recursion described earlier in Algorithm \ref{alg:AMS} can also be
used to find maximal irreducible sets for multi-source multi-sink
networks with correlated sources, replacing $\varphi(\cdot)$ by
$\mu_B(\cdot)$.}

\begin{example}[Butterfly network]\label{ex:Butterfly}
The maximal irreducible sets for the butterfly network in Figure \ref{butterfly} are
  \begin{multline}
    \{1,2\}, \{1,5\}, \{1,7\}, \{1,8\}, \{2,4\}, \{2,7\}, \{2,9\},
    \{3,4,5\}, \\
    \{3,4,8\}, \{3,7\}, \{3,8,9\}, \{4,5,6\}, \{5,6,9\}, \{6,7\},
    \{6,8,9\}.
  \end{multline}
\end{example}

{
 \begin{lemma}\label{lemma:3}
Consider a network FDG as defined in Definition \ref{def:FDG-B}.  
Suppose $h$ is a polymatroid on the ground set $(Y_{s}, U_{e}, s\in\sessions, e\in \edges)$, satisfying 
\eqref{eq:Networkcoding} and \eqref{eq:networkdecoding}. Then, one can extend $h$ to a polymatroid $h^{'}$ on the ground set  
$$
\X =\{ U_{e}, e\in\edges \} \cup
\{ Y_{s}, s\in\sessions\} \cup  \{ \hat Y_{s}^{i}, s\in\sessions, i \in b(s)\},
$$  
such that $h^{'}$ satisfies the network FDG.
\end{lemma}
\begin{IEEEproof}
The construction of $h^{'}$ is as follows. 
For any subset $\A$ of $\X$, let 
\[
\theta \triangleq 
\{
s\in\sessions : \: Y_{s} \not\in \A \text{ and }  \hat Y_{s}^{i}  \not\in \A, \: \forall i \in b(s)
\}
\]
and 
\[
\delta \triangleq 
\{
e \in\edges : \: U_{e}  \in \A  
\}
\]

Define 
\begin{align}
h^{'}(\A) =  h(U_{e}, \: e\in\delta,  Y_{s},\: s\not\in \theta).
\end{align}
It can then be verified directly that $h^{'}$ satisfies the network FDG, Definition \ref{def:FDG-B}.  
\end{IEEEproof}
}

We can now state our first main result, an easily computable outer bound for the capacity
region of a network coding system.

{
\begin{theorem}[Functional Dependence Bound]\label{thm:bound:b}
Consider a network coding problem and its induced network FDG $(\X, \edges^{*})$. 
If  $\B=\{U_{\A},Y_{\W^c}\}$ is a maximal irreducible set (with respect to Procedure B) in  $(\X, \edges^{*})$ and $(c_{e}, e\in\edges)$  is achievable, then 
\begin{align}
\sum_{e \in \A} c_e \geq  H\left(Y_{\W}\mid Y_{\W^c}\right). \label{eq:thm:bound:b}
\end{align}
In the special case when sources are independent, then inequality \eqref{eq:thm:bound:b} is reduced to 
\begin{align}
\sum_{e \in \A} c_e \geq  \sum_{s\in\W }H\left(Y_{s}\right).
\end{align}
\end{theorem}

\begin{IEEEproof}
Let 
$h$ be   a polymatroid  in $  \C_{1}\cap \C_2\cap\C_3 \cap\C_4$.
Then by Lemma \ref{lemma:3}, we can extend $h$ to a polymatroid $h^{'}$ over 
the ground set $\X$ satisfying the network FDG. 
Suppose $\B$ is a maximal irreducible set. Then
\[
h(\B) = h^{'}(\B) = h(Y_{s},  s\in\sessions). 
\]
Then by Theorem \ref{thm:1}, the result follows. 
 \end{IEEEproof}

Let $\boldsymbol{\M}_{B}$ be the set of all maximal irreducible set $\{U_{\A},Y_{\W^c}\}$ with respect to Procedures B and let
\begin{align}
  &\R_{FD} \nonumber\\
  &\triangleq 
  \bigcap_{\{U_{\A},Y_{\W^c}\} \in \boldsymbol{\M}_{B}}
  \left\{(c_{e}, e\in\edges) :
 \sum_{e \in \A} c_e \geq H\left(Y_{\W}| Y_{\W^c}\right)
 \right\}.\label{eq:R_FD}
\end{align}}
%%%%%%%%%%%%%%%%%%%

\begin{example}[Butterfly network]
  \label{ex:FDbd_butterfly}
  The functional dependence bound for the butterfly network of
  Figure~\ref{butterfly}, with correlated sources $Y_1$ and $Y_2$ is as
  follows (using the maximal irreducible sets in
  Example~\ref{ex:Butterfly}).
  \begin{align*}
    \{c_2, c_5, c_7\} &\geq h\left(Y_1\mid Y_2\right)  \\
    \{c_3, c_5, c_6\} &\geq h\left(Y_2\mid Y_1\right) \\
    \{c_1+c_5, c_4+c_5, c_1+c_2+c_3,& \\
    \quad\quad c_1+c_2+c_6, c_1+c_6+c_7,& \\
    \quad\quad c_2+c_3+c_4, c_2+c_3+c_4,& \\
    \quad\quad c_3+c_4+c_7,c_4+c_6+c_7\}&\geq h\left(Y_1,Y_2\right)
  \end{align*}
  If the sources $Y_1$ and $Y_2$ are instead independent, we obtain
  \begin{align}
    h(Y_1) &\leq \{c_2, c_5, c_7\} \label{eq:butterly rateR1}\\
    h(Y_2) &\leq \{c_3, c_5, c_6\} \label{eq:butterly rateR2}\\
    h(Y_1)+h(Y_2) &\leq \{c_1+c_5, c_4+c_5, c_i+c_j:\nonumber \\
    & \quad\quad  i \in \{2,5,7\}, j \in \{3,5,6\}\}  \label{eq:butterly rateR1+R2}
  \end{align}
  Note that the first two bounds $c_1+c_5, c_4+c_5$ on the sum rate
  in~\eqref{eq:butterly rateR1+R2} follow from
  the maximal irreducible sets $\{3,7\},\{6,7\}$ described in
  Example~\ref{ex:Butterfly}. The last nine bounds  are consequences of the individual
  rate bounds in~\eqref{eq:butterly rateR1} and~\eqref{eq:butterly
    rateR2}.
\end{example}

 \begin{remark}
 For single source multicast networks, the bound in Theorem~\ref{thm:bound:b}
 will be reduced to the max-flow bound~\cite[Theorem 11.3]{Yeu02} and hence is
  tight. {Summarizing  \eqref{eq:bounds} and Theorem \ref{thm:bound:b}, we have
%\begin{theorem}
  \begin{equation}
    \R  \subseteq \R (\bar\Gamma^*) \subseteq
    \R (\Gamma) \subseteq \R_{FD} .
  \end{equation}}
%\end{theorem}
The capacity region for the special case of multicast networks in
which all correlated sources are demanded by all sinks was established by
Han~\cite{Han11} using a simple cut-set based
characterization.  The cut-sets used by Han~\cite{Han11} are in fact the
maximal irreducible sets, yielding the following corollary.
\end{remark}

\begin{corollary}[When every  sink node demands all sources]\label{cor:AllsourceAllsinkCorrelated}
  For multicast networks in which all correlated sources are demanded
  by all sinks, $\R  = \R_{FD} .$
%  \begin{equation*}
%   \R^{cs} = \R_{FD}^{cs}.
% \end{equation*}
\end{corollary}

%%%%%%%%%%%%%%%%%%%%%%%%%%%%%%%%%%%%%
\subsection{When Sources are Independent}\label{sec:FDBdIS}

In this subsection, we further consider the special case when sources are independent.
Unlike the case when sources are correlated,
the problem of characterizing graphical bounds for networks with
independent sources has been well investigated~\cite{CovTho06,AhlCai00,Bor02,YanYan06,KraSav06,HarKle06}.
A source independence constraint may imply additional functional
dependencies beyond those implied by the network coding and decoding
constraints alone. These additional functional dependencies
may in turn be used to improve of our characterization of the set
of achievable rate region.

{
To understand the new bound, we first begin with a review of some basic graph concepts.  
}
The $d$-separation criterion~\cite{Pea88} is a tool to infer certain
conditional independence relationships amongst a set of random
variables where (some of) their local conditional independence
relations are represented by a Bayesian Network (directed acyclic
graph). It has also been shown that the $d$-separation criterion is
valid for finding certain conditional independence in cyclic
functional dependence graphs~\cite{Kra98} (see
Definition~\ref{def:d-separation}). The \fd-separation
criterion~\cite{Kra98} is an extension of $d$-separation finding
certain conditional independence relationships in FDGs. In this section, we generalize this result by
showing that \fd-separation can be used to find conditional
independence relationships for polymatroidal variables
represented by an FDG.

\def\An{{{An}}}
{
\begin{definition}[Ancestral graph]
Consider a directed graph $\graph^{*}=(\X,\edges^{*})$ induced by a network coding problem. For any subset $\A \subseteq \{ Y_{s}, s\in\sessions, U_{e}, e\in\edges  \}$, 
let $\An(\A)$ denote the set of all nodes in $\{ Y_{s}, s\in\sessions, U_{e}, e\in\edges  \}$ such that for every node $u \in \An(\A$), there is a directed path from $u$  to some node $v$ in  $\A$ in the subgraph $ \bar \graph^* \triangleq \graph^* \setminus \{e: {\eht{e}{Y_s}}, s \in \sessions \}$.

The ancestral graph  with respect to $\A$  (denoted by
  $\graph^{*}_{An(\A)}$)  is  a subgraph of $\graph^{*}$ consisting of nodes $\A \cup \mathrm{An}(\A)$ and
  edges $e \in \edges^{*}$ such that $\head{e},\tail{e} \in \A \cup
  \mathrm{An}(\A)$.

\end{definition}

%{\bf Terence: Can we use our own definition for fdg in the following two definitions?}
\begin{definition}[$d$-separation]\label{def:d-separation}
  
%  Let $\graph$ be a functional dependence graph according
%  to~\cite{Kra98}\footnote{We remind the reader that Kramer's
%    definition is a special case our Definitions
%    \ref{def:FDG} and~\ref{def:FDG-B}.}.  

    A set
  $\C$ $d$-separates $\A$ and $\B$ in a network FDG $\graph^*$ if the nodes in
  $\A$ and the nodes in $\B$ are disconnected in what remains of
  $\graph^* _{An(\A,\B,\C)}$ after removing all edges outgoing
  from nodes in $\C$.
%\end{enumerate}
\end{definition}

\begin{definition}[\fd-separation \cite{Kra98}]\label{def:fd-separation}
  Let $\graph^*$ be a network FDG.  A set $\C$ \fd-separates $\A$ and $\B$ in
  $\graph^*$ if the nodes in $\A$ and the nodes in $\B$ are disconnected
  in what remains of $\graph^* _{An(\A,\B,\C)}$ after removing
  all edges outgoing from nodes in $\C$ and subsequently,
  recursively removing all edges that have no source nodes as ancestors.
%\item Consider undirected version of the graph.
%\end{enumerate}
\end{definition}
}

Now we show that \fd-separation is valid for polymatroidal variables
represented by the subgraph $\bar{\graph^*}$ of network FDG
(Definition \ref{def:FDG-B}). First, note that the subgraph
$\bar{\graph^*}$ of network FDG is a functional dependence graph in
the sense of~\cite{Kra98} (with random variables replaced by
polymatroidal variables) since the vertices in $\bar{\graph^*}$ represent
source and edge variables, the edges in $\bar{\graph^*}$ represent
functional dependencies between the variables and the vertices
representing the source variables have no incoming edges.

%%%%%%%%%%%%%%%%%%%%%
\begin{lemma}\label{lem:pseudo_fd-sep}
  If the subset of nodes $\C$ \fd-separates $\A$ and $\B$ in the subgraph $\bar{\graph^*}$
  of a network FDG $\graph$ for $h$,  
  then $ I_{h}\left(\A ; \B \mid \C\right) =0$.
\end{lemma}
\begin{IEEEproof}
%\tc{Terence: Can you revise the proof?}
  By Definition~\ref{def:fd-separation}, $\varphi(\C)$ (see
  Definition~\ref{def:phi}) $d$-separates $\A$ and $\B$ in
  $\bar{\graph^*}$. But, $d$-separation is implied by the
  \emph{semi-graphoid} axioms (see~\cite[Chapter 3]{Pea88}) which are
  also satisfied by polymatroidal variables. Hence, if
  $\varphi(\C)$ $d$-separates $\A$ and $\B$ in $\bar{\graph^*}$
  then $I_h\left(\A ; \B\mid \varphi(\C)\right) =0$. By Lemma \ref{thm:fd}, $h\left(\varphi(\C)\right)=h(\C)$ and hence
  \begin{align*}
    h\left({\A\varphi(\C)}\right) + h\left({\B\varphi(\C)}\right) - h\left({\varphi(\C)}\right) - h\left({\A\B\varphi(\C)}\right)  = 0 % \qquad \qquad \\
\end{align*}
implies
  \begin{align*}
I_h\left(\A;\B\mid \C\right) = h\left({\A\C}\right) +
    h\left({\B\C}\right) - h\left({\C}\right) - h\left({\A\B\C}\right)
     = 0.
\end{align*}
\end{IEEEproof}

%\subsubsection{A Graphical Bound for Independent Sources}
In the following, we will give a tighter graphical bound for networks when  sources are independent. We will follow  a similar approach used to derive   
Theorem~\ref{thm:mainresult1} by finding  maximal irreducible sets induced by
\fd-separation in subgraph $\bar\graph^*$ of network FDG.

{
\begin{definition}[Procedure C]\label{def:fd1}
 Consider a network FDG as defined in Definition \ref{def:FDG-B}.  
  For any sets $\A,\B\subseteq\X$, we say $\A$ determines $\B$
(with respect to  Procedure C) if there are no elements
  of $\B$ remaining after the following procedure:

%  For $\B\subseteq \A \subseteq \nodes$, $\A$ determines $\B$. For disjoint sets $\A,\B\subseteq\nodes$ we say $\A$ determines $\B$
%  (denoted $\A\twoheadrightarrow\B$) in the network FDG $\graph=(\nodes,\edges)$ of a given multi-source
%  multi-sink network with independent sources, if there are no
%  elements of $\B$ remaining in the resulting graph $\graph^*$ after
%  the following procedure:

{\bf Procedure C: }

\begin{quote}

 \begin{enumerate}
 \item Remove all edges outgoing from $\A$ and subsequently
   recursively remove all nodes and edges with no incoming edges and
   nodes respectively and all nodes and edges with no source nodes as
   ancestors. Call the resulting graph $\tilde\graph^*$.
 \item If there exists any $Y_s$ disconnected from any $\hat Y^i_s$ in
   $\tilde \graph^*_{An(Y_s,\hat Y^i_s, \A)}$ then from $\tilde\graph^*$ (a)
   remove $\hat Y^i_s$ for all $i \in \{1,...,|b(s)|\}$ and (b)
   subsequently recursively remove all nodes and edges with no
   incoming edges and nodes respectively. Call the resulting graph
   $\tilde\graph^*$, go to Step 2. Else terminate.
  \end{enumerate}
\end{quote}
We will use $\A \twoheadrightarrow \B$ to denote that  $\A$ determines $\B$
 with respect to  Procedure C. 
\end{definition}

Note that Step 2 of Definition \ref{def:fd1} uses \fd-separation. 
The concepts for blanket, irreducibility are similarly defined with respect to Procedure C.
Specifically, for a given set $\A$, 
its blanket (with respect to Procedure C) is  denoted by  
 $\psi(\A)$ and is defined as  the largest set of nodes with
$\A\twoheadrightarrow \psi(\A)$.
  A set of nodes $\B$  is called
  \emph{irreducible} (with respect to Procedure C) if there is no $\A\subseteq\B$ such that
  $\A \twoheadrightarrow \B$. An irreducible set $\A$ is \emph{maximal}  if
  $\psi(\A)=\X$.
 In addition, if $\A$ and $\B$ are maximal irreducible sets, then 
\[
h(\A) = h(\B).
\]
Furthermore, the recursion described earlier in Algorithm \ref{alg:AMS} can now be
used to find maximal irreducible sets for multi-source multi-sink
networks with independent sources, replacing $\varphi(\cdot)$ by
$\psi(\cdot)$.

}

\begin{corollary}\label{cor:phi sub psi}
  For any given network FDG $\graph^{*}
  =(\X,\edges^{*})$, 
  \begin{equation*}
    \mu_B(\A) \subseteq \psi(\A), \forall \A \subseteq \X.
  \end{equation*}
\end{corollary}

We remark that, there may exist some $\A \subseteq \nodes$ in $\graph
=(\nodes,\edges)$ such that $\mu_B(\A) \subsetneq \psi(\A)$ (refer to
Example \ref{ex:ButterflyIS} in which $\mu_B(\{4,5\}) \subsetneq
\psi(\{4,5\})=\nodes$).

{
\begin{lemma}\label{thm:fd1}
%Suppose $\A \rightarrowtail \B$ in the network FDG and $h$ satisfies 
%\eqref{eq:Networkcoding} and \eqref{eq:networkdecoding}.
If $\A, \B \subseteq \{ Y_{s}, s\in\sessions, U_{e}, e\in\edges  \}$ and 
$\A  \twoheadrightarrow \B$ in a network FDG $\graph^*$, then $h(\B \mid \A) = 0$.  
\end{lemma}
}
\begin{IEEEproof}
%\tc{Can you edit the proof accordingly?}
  Suppose $\A \twoheadrightarrow \B$. Let $\U$ be the set of
  variables removed by Step 1 in Definition~\ref{def:fd1}. Then
  by Lemma \ref{thm:fd}, $h\left(\U\mid\A\right)=0$.
Now, let $\Y$ be the set of all nodes representing the estimates
  $Y^{i}_s, i \in b(s), s \in \W \subseteq \sessions$ removed by
  Step~2(a) in Definition~\ref{def:fd1}. Then, by the definition of
  \fd-separation in the subgraph $\bar \graph^*$
  and Lemma~\ref{lem:pseudo_fd-sep},  $I_h(Y_s;\hat Y_s^i\mid\A)=0, s \in \W, i \in b(s)$.
  But the decoding constraints~\eqref{eq:networkdecoding} imply
  $\hat Y_s^i=Y_s$ then for $s\in \W$,
  \begin{equation*}
    h(Y_s\mid\A)=h(\hat Y_s^i\mid\A)=0 \Rightarrow h(\W\mid\A)=0.
  \end{equation*}
  Let $\Z$ be the set of all variables removed by Step~2(b) in
  Definition \ref{def:fd1}. Then by Lemma \ref{thm:fd},
  \begin{equation*}
    h(\Z\mid\A)=0.
  \end{equation*}
  Since $\psi(\A)=\U \cup \Y \cup \W \cup \Z$,
  \begin{equation*}
    \A \twoheadrightarrow \B \Rightarrow \B \subseteq \U \cup \W \cup \Z
  \end{equation*}
  and hence
  \begin{equation*}
    h(\U \W \Z\mid\A)=0 \Rightarrow h(\B\mid\A)=0.
  \end{equation*}
\end{IEEEproof}

\begin{figure}[htbp]
  \centering
  \includegraphics[scale=.9]{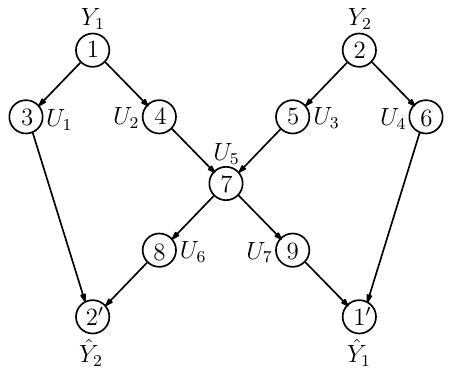}
  \caption{Subgraph $\bar\graph^*$ of network FDG for the butterfly network.}
  \label{fig:gbar}
\end{figure}

\begin{example}[Butterfly Network, Independent Sources]
  \label{ex:ButterflyIS}
  Figure~\ref{fig:gbar} shows the subgraph $\bar\graph$ for
  network FDG (Figure~\ref{mul2}) of the butterfly network
  (Figure~\ref{butterfly}).  The independent source maximal
  irreducible sets are
  \begin{multline*} \{1,2\}, \{1,5\}, \{1,7\}, \{1,8\},
    \{2,4\}, \{2,7\}, \{2,9\},\{3,7\},\\
    \{4,5\},\{4,7\},\{4,8\},\{5,7\},\{5,9\},\{6,7\},\{3,8,9\},\{6,8,9\}.
  \end{multline*}
  The sets $\{4,5\}$, $\{4,7\}$, $\{4,8\}$, $\{5,7\}$, $\{5,9\}$ are
  new maximal irreducible sets found by replacing $\varphi(\cdot)$ by
  $\psi(\cdot)$ in Algorithm~\ref{alg:AMS}. Source independence is an
  essential ingredient to find these new maximal irreducible sets.
  However independence is not necessary to find the other maximal
  irreducible sets. Also note that the maximal irreducible sets $\{3,4,5\}$,
  $\{4,5,6\}$, $\{3,4,8\}$, $\{5,6,9\}$ previously found by
  Algorithm~\ref{alg:AMS} with $\varphi(\cdot)$ are further reduced to
  $\{4,5\}$, $\{4,8\}$, $\{5,9\}$ using source
  independence via $\psi(\cdot)$.
\end{example}

{
It may be of theoretical interest to know which functional
dependencies are implied by local encoding/decoding functions and
which involve source independence. Algorithm \ref{alg:AMS} with
$\mu_B(\cdot)$ and $\psi(\cdot)$ can be used to answer this question.
Our main result for independent sources is as follows (the proof is
similar to Theorem~\ref{thm:mainresult1}).
}
\begin{theorem}[Functional dependence bound, independence contraints]\label{thm:mainresult2}
Consider a network coding problem with independent sources and its induced network FDG $(\X, \edges^{*})$. 
If  $\B=\{U_{\A},Y_{\W^c}\}$ is a maximal irreducible set (with respect to Procedure C) in  $(\X, \edges^{*})$ and $(c_{e}, e\in\edges)$  is achievable, then 
\[
\sum_{e \in \A} c_e \geq  \sum_{s\in\W }H\left(Y_{s}\right).
\]

\end{theorem}

Let $\boldsymbol{\M}_{C}$ be the set of all maximal irreducible set $\{U_{\A},Y_{\W^c}\}$ with respect to Procedure C  and let 
\begin{align}
  &\R_{FD}^{\perp} \nonumber\\
  &\triangleq 
  \bigcap_{\{U_{\A},Y_{\W^c}\} \in \boldsymbol{\M}_{C}}
  \left\{(c_{e}, e\in\edges) :
 \sum_{e \in \A} c_e \geq  \sum_{s\in\W }H\left(Y_{s}\right)
 \right\}.\label{eq:R_FDI}
\end{align}

\begin{corollary}\label{thm:RidsubsetRfd}
%  For any point-to-point network with error-free links
When sources are independent,
  \begin{equation}\label{eq:RidsubsetRfd}
    \R_{FD}^{\perp} \subseteq \R_{FD}
  \end{equation}
  and there exists a network for which the inclusion is strict.
\end{corollary}
%\begin{IEEEproof}

  $\R_{FD}^{\perp} \subseteq \R_{FD}$ follows from Corollary~\ref{cor:phi sub
    psi} and Theorems~\ref{thm:mainresult1}
  and~\ref{thm:mainresult2}. Strict inclusion is demonstrated in
  Example~\ref{ex:nspde} in Section~\ref{sec: comparison}.
%\end{IEEEproof}

%%%%%%%%%%%%%%%%%%%%%%%%%%%%%%%%%
\section{Comparison}\label{sec: comparison}
We now compare our bounds $\R_{FD}$ and $\R_{FD}^{\perp}$ with some known bounds. It should be noted that a
comparison of all these known bounds does not seem to have been previously
performed in the literature. This is in part due to the different
forms of the bounds. In contrast, our unifying framework enables us to
complete this comparison. In addition to establishing the comparative strength of the bounds, 
the comparison may provide insight into the essential
technical ingredients for  characterization of the bounds and hence
helps answer why one bound is better than (or similar to) another. For comparison purpose we assume that $a(s)$ are singletons for all $s \in \sessions$.

%%%%%%%%%%%%%%%%%%%%%%%%%%%
\subsection{Cut-Set Bound}
%\subsubsection{Cut-Set Bound}\label{sec:Cut-Set Bound}
The cut-set bound~\cite[Theorem 15.10.1]{CovTho06} is an outer bound
on the capacity region of general multi-terminal communication
networks.  For a subset of sessions $\W\subseteq\sessions$, let
$\boldsymbol{\T}_{\W} = \{\T_{\W}\subseteq\nodes:
a(s)\in\T,b(s)\cap\T^c\neq\emptyset, \forall s\in\W\}$ be the collection of
all subsets of nodes $\T_{\W}$ such that these source sessions are
available to nodes in $\T_{\W}$, and at least one node in the complement
$\T_{\W}^c$ demands each session.  Further define
$\edges(\T_{\W})=\{e\in\edges:\tail{e}\in\T_{\W},\head{e}\in\T_{\W}^c\}$ as the
cutset of edges separating $\T_{\W}$ and $\T_{\W}^c$.  For our case of interest,
networks consist of error free point-to-point links, and the cut-set
bound reduces to the following simple upper bound~\cite{Bor02}, which
is identical to the max-flow bound of~\cite{AhlCai00} (see
also~\cite{Yeu08}).
\begin{theorem}\label{lem:cut-set bound}
  For a network of error free point-to-point channels, if $c_e,e\in \mathcal E$ is achievable, then
  \begin{equation}
    \label{eq:cutsetbound}
    \sum_{s\in\W} H(Y_s) \leq \sum_{e\in\edges(\T_{\W})} c_e.
  \end{equation}
\end{theorem}
%The inequalities defining the cut-set bound together with
%non-negativity of entropy characterize a \emph{cut-set region} in the
%non-negative orthant $\Real_{+}^{|\sessions|}$ of
%$|\sessions|$-dimensional Euclidean space.
{
Define the corresponding outer bound region,
\begin{align}
  &\R_{CS} \nonumber \\
  &\triangleq \bigcap_{\W\subseteq \sessions,\edges(\T_{\W})} \left\{
(c_e:e \in \edges): \sum_{s\in\W}
H(Y_s) \leq \sum_{e\in\edges(\T_{\W})} c_e
\right\}.\label{eq:cutsetregion}
\end{align}
}

Now, we compare the cut-set bound with functional
dependence bound (Theorem~\ref{thm:mainresult1}). In the proof of the
cut-set bound~\cite[Theorem 15.10.1]{CovTho06} the decoding
constraints are only loosely enforced. The source messages $y_s: s
\in\W \subseteq \sessions$ transmitted from nodes in $\mathcal T_{\W}$ to
nodes in $\mathcal T_{\W}^c$ can be decoded from symbols received at nodes
in $\mathcal T_{\W}^c$ and other source messages $y_s: s \in\W^c$, i.e.,
$Y_{\W} = f(\{U_{e}: \head{e} \in \mathcal T_{\W}, \tail{e} \in \mathcal
T_{\W}^{c}\}, Y_{\W^c})$. This is a kind of joint decoding, potentially
with extra side information, and hence does not enforce the decoding
constraints independently at each sink (this will be clear from
Example~\ref{ex:cut-setbutterfly}). {To simplify notations we consider unicast network.}
%Below we first prove that the
%cut-set bound cannot be tighter than the functional dependence bound
%and then demonstrate that the functional dependence bound is strictly
%tighter than the cut-set bound using the two source butterfly network.
\begin{theorem}\label{thm:RfdsubsetRcs}
  %For any point-to-point network  with error-free links
  %\begin{equation}\label{eq:RfdsubsetRcs}
    $\R_{FD} \subseteq \R_{CS}$
  %\end{equation}
%  where $\R_{FD}$ is the functional dependence region~\eqref{eq:R_FD}
 % of Theorem~\ref{thm:mainresult1}
%  and $\R_{CS}$ is the cut-set region~\eqref{eq:cutsetregion} 
and the
  inclusion can be strict.
\end{theorem}
\begin{IEEEproof}
  For $Y_{\W}$ available at some nodes in $\T_{\W} \subseteq \nodes$, let
  $\A = \{e: \head{e} \in \mathcal T_{\W}, \tail{e} \in \mathcal T_{\W}^{c}\}$
  be any cut-set defining $\R_{CS}$. Then to prove $\R_{FD} \subseteq
  \R_{CS}$ it is sufficient to prove that in the network FDG
  $Y_{\W} \subseteq \mu_B (\A, Y_{\W^c})$. %Now, n
  {Consider paths from $Y_s,s \in \sessions$ to $\hat{Y_s}$.} Note that in the network FDG of the given
  network, every path from nodes (representing sources of) $Y_s:s \in
  \W$ to nodes (representing sinks) $\hat{Y}_s : s \in {\W}$ in
  $\mathcal T_{\W}^c$ intersects some nodes (representing the edges) in
  $\A$.
  Then by Definitions \ref{def:fd} and \ref{def:phi},
  \begin{equation}\label{eq:RfdsubsetRcs1}
    Y_{\W} \subseteq \mu_B (\A, Y_{\W^c}).
  \end{equation}
  This is because other paths to sink nodes $\hat{Y}_s : s \in {\W}$
  in $\mathcal T_{\W}^c$ can only be from nodes $Y_s: s \in {\W^c}$. Hence
  there are no other paths {from $Y_s,s \in \sessions$} to sink nodes of $Y_s:s \in {\W}$ in
  $\mathcal T_{\W}^c$ except those intersecting $\A$ and those containing
  nodes in $Y_s: s\in {\W^c}$. By~\eqref{eq:RfdsubsetRcs1} and
  Theorem~\ref{thm:mainresult1}, $\R_{FD} \subseteq \R_{CS}$. Example~\ref{ex:cut-setbutterfly} below shows $\R_{FD} \subsetneq
  \R_{CS}$ for the butterfly network.
\end{IEEEproof}

{
\begin{example}\label{ex:cut-setbutterfly}
  For the butterfly network of Figure~\ref{butterfly}, the functional
  dependence bound of Theorem~\ref{thm:mainresult1}, is strictly
  tighter than the cut-set bound, Theorem \ref{lem:cut-set bound}. More
  specifically, the cut-set bound is
  \begin{align*}
    H(Y_1) &\leq \{c_2, c_5, c_7\}\\
    H(Y_2) &\leq \{c_3, c_5, c_6\}\\
    H(Y_1)+H(Y_2) &\leq \{c_1+c_4+c_5, c_i+c_j:\nonumber \\
    & \quad\quad  i \in \{2,5,7\}, j \in \{3,5,6\}\}
  \end{align*}
 % Now, assign the edge capacities $c_e=1, e \in\{1,4\}$ and $c_e=3, e \in\{2,3,5,6,7,8\}$.
  %Then the cut-set sum-rate bound evaluates to
  %\begin{align}
   % H(Y_1)+H(Y_2) \leq c_1+ c_4+ c_5 =5.
 % \end{align}
  On the other hand, \eqref{eq:aaa} is tighter using the functional dependence bound (via
  the maximal irreducible sets $\{3,7\}$ and $\{6,7\}$ corresponding
  to the sets of variables $\{U_{1},U_{5}\}$ and
  $\{U_{4},U_{5}\}$ respectively, see Examples \ref{ex:Butterfly} and
  \ref{ex:FDbd_butterfly}).
  \begin{align}\label{eq:aaa}
    H(Y_1)+H(Y_2) &\leq  \{c_1+ c_5, c_4+ c_5\}
  \end{align}
\end{example}
}

%%%%%%%%%%%%%%%%%%%%%%%%%%%%%%
\subsection{Network Sharing Bound}
The network sharing bound~\cite[Theorem 1]{YanYan06} is defined for a
special type of multiple unicast (each session is demanded at only one
sink) networks called $|\sessions|$-pairs three-layer
networks~\cite{YanYan06} where $\sessions$ is the set of source
sessions.  Three-layer networks are a network extension of the
distributed source coding model of~\cite{YeuZha99}. Each channel $e
\in \edges$ of finite capacity $c_e$ has direct access to certain
source sessions and each sink has access to certain channels.

{
\begin{definition}\label{def:three-layer network}
  A \emph{three-layer network} is a directed  
  graph $\graph=(\nodes, \edges)$ characterized by a tuple $(\edges^{'}, \alpha, \beta)$ such that $|a(s)|=|b(s)|=1$ for all $s\in\sessions$. Here, 
  \begin{enumerate}
  \item $\edges^{'}$    is  the set of edges in the middle layer such that   $\tail e \neq \head{f}$ for all distinct $e,f \in \edges^{'}$. In other words, all the middle layer edges are not directly connected.

 \item  
  source connection $\alpha : \edges^{'} \mapsto  2^{\sessions}$ specifies the first layer edges, which have  the form 
 $(a(s) , \tail{e})$ for $s\in \alpha(e)$.

  \item 
 sink connection $\beta : \edges^{'} \mapsto  2^{\sessions}$ specifies the third layer edges, which have  the form
 $(\head{e}, b(s))$ where $s \in \beta(e)$.

  \end{enumerate}
    \end{definition}

To define the network sharing bound, we assume without loss of generality that $\sessions$ is a {strict} totally ordered set (with the binary order relation  $\prec$). {We say $s_{i(\prec)}=k$ if, given the total order $\prec$, $s_i$ is the $k$th element in the set with respect to $\prec$.}

\begin{definition}\label{def:ns edge-set}
For a given three-layer network (see Definition \ref{def:three-layer network}),
a \emph{network sharing edge-set} $\mathcal F$   {with respect to $\prec$ on a subset} of sources $\W \subseteq \sessions$    is the set 
  \begin{equation*}
    \mathcal F (\W, \prec) \defined \{e: \beta(e) \cap \W \neq \emptyset, \alpha(e)
    \nsubseteq \W[{\beta(e)}] \}
  \end{equation*}
  where    
  $\W[{\beta(e)}] \defined \{s_{i} \in
  \W:  s_i \prec  s_j , s_j \in \beta(e)\}$.
In other words,  $\mathcal F(\W, \prec)$ may be viewed as the set of edges $e$ such that there exists $s \in \alpha(e)$ and $s' \in \beta(e)$ satisfying $ s'  \prec  s  $. 
%The edges in a network
%  sharing edge-set carry the set of variables $U_{\mathcal F}$.
\end{definition}
}

\begin{example}[$2$-pairs three-layer butterfly network]
  Figure \ref{fig:3-layerButterfly1} shows an example of a three-layer
  network, where $\sessions=\{s_1,s_2\}$ the sources are located at
  nodes $a(s_1)=1$, $a(s_2)=2$ and demanded at nodes $b(s_1)=9$,
  $b(s_2)= 10$. The source and sink connections (shown with dashed
  edges) are
%  \begin{align*}
    $\alpha(e_1)=\{s_1\}$, $\beta(e_1)=\{s_2\}$, $\alpha(e_2)=\{s_1,s_2\}$, $\beta(e_2)=\{s_1,s_2\}$, $\alpha(e_3)=\{s_2\}$ and $\beta(e_3)=\{s_1\}$.
%  \end{align*}
  \begin{figure}[htbp]
    \centering
    \includegraphics[scale=.9]{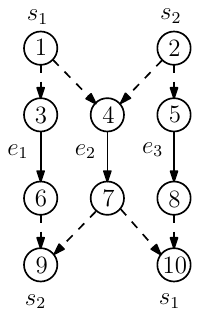}
    \caption{$2$-pairs three-layer butterfly network.}\label{fig:3-layerButterfly1}
  \end{figure}
\end{example}
%\begin{definition}

{
\begin{theorem}[Theorem 1, \cite{YanYan06}]\label{thm:NS bound}
Consider a three-layer unicast network $(\nodes, \edges)$. 
If the edge capacity tuple $(c_{e}, e\in\edges)$ is achievable, then   
\begin{equation}\label{eq:nsbound}
\sum_{s \in \mathcal{W}} H(Y_s) \leq \sum_{ e \in \mathcal F(\W, \prec)}  c_{e}.
\end{equation}
%where ${\sigma}$ is a permutation mapping of source sessions in
%$\mathcal W \subseteq \sessions$ and $\mathcal W[\beta(e)] \defined
%\{s_{i} \in \mathcal W: \sigma(s_i) < \sigma(s_j), s_j \in
%\beta(e)\}$.
%Corollary, by optimising the choice of the total order $\prec$,

\end{theorem}
}

Similar to~\eqref{eq:cutsetregion},  define the \emph{network sharing
  region} as the subset of $\Real_{+}^{|\edges|}$ such that~\eqref{eq:nsbound} holds.
\begin{equation}\label{eq:nsregion}
\R_{NS}  \triangleq \bigcap_{\mathcal W \subseteq \sessions, \text{ and total order } \prec}
\left\{ (c_e:e \in \edges): \text{~\eqref{eq:nsbound} holds} \right\}.
\end{equation}

Now, we show that the functional dependence bound,
Theorem~\ref{thm:mainresult1}, and the network sharing bound
\cite{YanYan06}, Theorem~\ref{thm:NS bound}, are identical
(restricting attention to three-layer networks). This also proves that
Theorem~\ref{thm:mainresult2} (for independent sources) is better than
the network sharing bound. %, since we have already shown it to be better
%than the functional dependence bound.

\begin{theorem}\label{thm:nsequalfd}
For a three-layer network,
%\begin{equation}
$$\R_{NS}= \R_{FD}.$$
%\end{equation}
%where $\R_{NS}$ is the network sharing region \eqref{eq:nsregion} and $\R_{FD}$ the functional dependence region \eqref{eq:R_FD}.
\end{theorem}

The proof of the theorem follows from Lemmas \ref{lem:NSdirect} and
\ref{lem:NSconverse} below. 
Also note that the
proof of the network sharing bound uses subadditivity of entropies
similar to Theorem~\ref{thm:mainresult1}. For simplicity and clarity, we prove Lemmas~\ref{lem:NSdirect}
and~\ref{lem:NSconverse} for $\W = \sessions$. With similar proof
methods, the following statements can be proved.
\begin{enumerate}
\item For any $\W \subseteq \sessions$ and some order
  $\prec$, let $U_{\mathcal F(\W,\prec)}$ be a set of variables flowing
  through network sharing edge-set $\mathcal F(\W,\prec)$
  then,
%  \begin{equation*}
    $\exists\mathcal B \in \boldsymbol{\M} : \mathcal B  \subseteq \mathcal F(\W,\prec) \cup \W^{c}$
%  \end{equation*}
  where $\boldsymbol{\M}$ is the collection of all maximal irreducible
  sets of the three layer network.% and
%  \begin{equation*}
 %   \mathcal F(\W) \defined \{e: \beta(e)\cap \W \neq \emptyset,
%    \alpha(e) \cap \W \nsubseteq \W[{\beta(e)}]\}.
%  \end{equation*}
\item For every maximal irreducible set of a given three-layer network
  there exists an equivalent set $\mathcal F(\W,\prec) \cup \W^{c}$ obtained
  by some ordering {(relation)} $\prec$.
\end{enumerate}

{
\begin{lemma}\label{lem:NSdirect}
  In a three-layer network, every network sharing edge-set contains a
  maximal irreducible set not containing any source variables.
  % in the bipartite FDG of the three-layer network.
  That is, for any network sharing edge-set $\mathcal F(\sessions, \prec)$,
  \begin{equation*}
    \exists\mathcal B \in \boldsymbol{\M} : \mathcal B  \subseteq U_{\mathcal F(\sessions, \prec)}
  \end{equation*}
  where $ \boldsymbol{\M}$ is the collection of all maximal
  irreducible sets in the FDG of the
  three-layer network.
\end{lemma}
}
\begin{IEEEproof}
  Let $U_{\mathcal F(\sessions, \prec)}$ be the set of network sharing edge-set
  variables obtained via order $\prec$. Then by the
  definition of the network sharing bound, Theorem~\ref{thm:NS bound},
  \begin{equation*}
    \{e :s_i \in \beta(e), {s_{i(\prec)}=1}\} \subseteq \mathcal F(\sessions, \prec)
  \end{equation*}
  and so
  \begin{equation*}
    Y_{\{s_i:s_{i(\prec)}=1\}} \in \varphi(U_{\mathcal F(\sessions, \prec)}).\footnote{We consider $\varphi(\cdot)$ since the network is unicast which yeilds cyclic FDG described in Section \ref{sec:FDG}.}
  \end{equation*}
  Also note that
  \begin{multline*}
    \{e : s_j \in \beta(e),s_{j(\prec)}=2\} \\
    \subseteq \mathcal F(\sessions, \prec) \cup \{
    U_e : s_i \in \alpha(e), s_{i(\prec)}=1\}
  \end{multline*}
  and so
  \begin{equation*}
    Y_{\{s_j:s_{j(\prec)}=2 \}} \in \varphi(U_{\mathcal F(\sessions, \prec)},
    Y_{\{s_i:s_{i(\prec)}=1\}}) = \varphi(U_{\mathcal F(\sessions, \prec)}).
  \end{equation*}
  In general,
  \begin{equation*}
    \{e : s_j \in \beta(e)\} \subseteq \mathcal F(\sessions, \prec) \cup \{e : s_i \in
    \alpha(e), s_{i(\prec)} <s_{j(\prec)}\}
  \end{equation*}
implies
  \begin{equation*}
    Y_{s_j} \in \varphi(U_\mathcal F(\sessions, \prec),Y_{\{s_i :
      s_{i(\prec)}<s_{j(\prec)}\}}) = \varphi(U_{\mathcal F(\sessions, \prec)}).
  \end{equation*}
  Therefore, for any network sharing edge-set under some order of source
  nodes induced by the relation $\prec$,
  \begin{equation*}
    \varphi (U_{\mathcal F(\sessions, \prec)})= U_\edges \cup Y_{\sessions}
  \end{equation*}
  and hence there exists $\mathcal B \subseteq U_{\mathcal F(\sessions, \prec)}$.
%  \begin{equation*}
%    \exists \mathcal B \subseteq U_\mathcal F.
%  \end{equation*}
\end{IEEEproof}

We remark that there could exist maximal irreducible sets which are
proper subsets of network sharing edge-sets. On the other hand, the
following lemma proves that, for a given maximal irreducible set, we
can always find an equivalent network sharing edge-set. The lemma also
describes the ordering induced by $\prec$ for which
$\mathcal B = U_{\mathcal F(\sessions, \prec)}$.

{
\begin{lemma}\label{lem:NSconverse}
  For any maximal irreducible set $U_{\mathcal F(\sessions, \prec)}$ not containing any
  source variables in the FDG of a given
  three-layer network, there exists the network sharing edge-set
  $\mathcal F(\sessions, \prec)$ obtained via a reordering of the source nodes.
\end{lemma}
}
\begin{IEEEproof}
  Let $\mathcal B=U_{\edges_1}, \edges_1 \subseteq \edges$ be a
  maximal irreducible set (not containing any source nodes). By
  Definition~\ref{def:IrredSet1}, $\varphi(U_{\edges_1}) =
  U_\edges \cup Y_\sessions$. Now, (recalling $\parent$ to be the set
  of parents~\eqref{eq:parent}) let
%  \begin{equation*}
    $\sessions_1\triangleq \{s_i: \pi(Y_{s_i}) \subseteq U_{\edges_1}\} \subseteq \sessions$
%  \end{equation*}
  be a set of sources which are immediate children of nodes in
  $U_{\edges_1}$ and are not children of any other nodes.  Also define
%  \begin{equation*}
    $\edges_2 \triangleq \{e: \pi(U_e) \subseteq Y_{\sessions_1}, e \not\in \edges_1\}.$
%  \end{equation*}
  Recursively define the following sets
  \begin{align*}
    \sessions_i\triangleq \Bigg\{s: \pi(Y_s) \subseteq \bigcup_{j \in
      \{1,...,i\}} U_{\edges_j}, s \not\in \bigcup_{j \in
      \{1,...,i-1\}} \sessions_j\Bigg\}\\
    \edges_i \triangleq \Bigg\{e: \pi(U_e) \subseteq \bigcup_{j \in
      \{1,...,i-1\}} Y_{\sessions_{j}}, e \not\in \bigcup_{j \in
      \{1,...,i-1\}} \edges_j\Bigg\}.
  \end{align*}
  Note that the nodes in $Y_{\sessions_i}$ have incoming edges only
  from edges in $U_{\edges_j}, j\leq i$ and nodes in $U_{\edges_i}$
  have incoming edges only from nodes in $Y_{\sessions_j}, j\leq
  i-1$. Also, for any $i \neq j$, $Y_{\sessions_i}$ and
  $Y_{\sessions_j}$, and $U_{\mathcal B_i}$ and $U_{\mathcal B_j}$ are
  disjoint. In the three-layer network,
  \begin{align*}
    \edges_i =& \Big\{e : \alpha(e) \subseteq \{\sessions_j : j \leq
    i-1\} = \sessions[{\beta(e)}],\\
    & \text{ \ \  \ \ \ \ \ \ \ \ \ \ \ \ \ \ \ \ \ \ \ \  \ \ \ \ \ \ \ \ \ \ \ \ \ \ \ \ } e \not\in \bigcup_{j \in \{1,...,i-1\}} \edges_j\Big\}
  \end{align*}
  where $i > 1$. But, by definition of the network sharing bound, edges $e \in
  \edges_i, i>1$ will not be included in the network sharing edge-set
  for any order relation $\prec$ such that $\{s_{i(\prec)}:s_i \in \sessions_1\}
  < ... < \{s_{i(\prec)}:s_i \in \sessions_m\}, \max_{s_{i} \in \sessions_m}
  s_{i(\prec)} = |\sessions|$ (ordering of sessions within each
  $\sessions_i$ is irrelevant).  Hence there exists a network sharing
  edge-set $\mathcal F (\sessions,\prec) \subseteq \edges_1$ such that $U_{\mathcal F(\sessions,\prec) }
  \subseteq U_{\edges_1}, U_{\edges_1} \in \boldsymbol{\M}$.
\end{IEEEproof}

\begin{remark}
Although the network sharing bound turns out to be the same as the
functional dependence bound for in a three-layer network, the functional
dependence bound is not a simple extension of the network sharing
bound for more general networks. In fact, the functional dependence
bound uses a completely different approach for characterizing
bottlenecks such that, given the network coding and decoding
constraints, variables flowing in a bottleneck determine all other
variables. Also, the network sharing bound is computationally more
complex compared to the functional dependence bound in the sense that all
possible orderings of the sources need to be considered to find a
network sharing edge-set. 
\end{remark}

%Corollary~\ref{thm:RfdsubsetRns} below follows from
%Corollary~\ref{thm:RidsubsetRfd} and Theorem~\ref{thm:nsequalfd}.
\begin{corollary}\label{thm:RfdsubsetRns}
  For a three-layer network
  \begin{equation}\label{eq:RfdsubsetRns}
    \R_{FD}^{\perp} \subseteq \R_{NS}
  \end{equation}
  where $\R_{FD}^{\perp}$ is the bound~\eqref{eq:R_FDI} for independent sources and $\R_{NS}$ the network sharing
  region~\eqref{eq:nsregion}. There exists a network for which $\R_{FD}^{\perp} \subsetneq \R_{NS}$.
\end{corollary}

\begin{IEEEproof}
A direct consequence of Corollary~\ref{thm:RidsubsetRfd} and Theorem~\ref{thm:nsequalfd}. %Terence: How to show that the inclusion is strict?{the example is mentioned in Corollary 4.}
\end{IEEEproof}

An important implication of Theorem~\ref{thm:nsequalfd} together with Theorem~\ref{thm:mainresult1} is that (1) the
network sharing bound can be applied to three-layer networks with
correlated sources and (2) the source independence constraint
is not exploited to characterize network sharing edge-sets.

%%%%%%%%%%%%%%%%%%%%%%%%%%%%%%%%
\subsection{Information Dominance and a New Bound}\label{sec:ID Comp}
The notion of information dominance and its graphical characterization
was introduced in \cite{HarKle06}.
\begin{definition}\label{def:ID}
  Given $\mathcal G=(\nodes,\edges)$, an edge set $\A \subseteq
  \edges$ \emph{informationally dominates} $\B \subseteq \edges$ if
  for all network codes $\tilde\phi$~\eqref{eq:globalcode} and
  $|\sessions|$-tuples of messages
  $\mathbf{x}=(x_1,...,x_{|\sessions|})$ and
  $\mathbf{y}=(y_1,...,y_{|\sessions|})$,
  \begin{equation*}
    \tilde{\phi}_{\mathcal
      A}(\mathbf{x}) = \tilde{\phi}_{\mathcal A}(\mathbf{y}) \implies
    \tilde{\phi}_{\mathcal B}(\mathbf{x}) = \tilde{\phi}_{\mathcal
      B}(\mathbf{y}).
  \end{equation*}
 Also define
  \begin{equation}
    \Dom(\A) \defined \{e: \A \text{ informationally dominates } e\}.
  \end{equation}
\end{definition}

\begin{definition}[$\mathcal G (\Dom(\A),s)$]\label{def:DomModifiedG}
  Given a graph $\graph=(\nodes,\edges)$, an edge set
  $\A\subseteq\edges$ and a source session $s\in\sessions$, $\graph
  (\Dom(\A),s)$ is the graph obtained by the following manipulation:
  \begin{itemize}
  \item remove edges and nodes that do not have a path to $\hat Y_s$ in $\graph$,
  \item remove all edges in $\A$,
  \item remove edges and nodes that are not reachable from a source edge in the remaining graph.
  \end{itemize}
\end{definition}
The conditions of the theorem below characterize $\Dom(\A)$.
\begin{theorem}[\cite{HarKle06}, Theorem 10]\label{thm:ID}
  For an edge set $\A\subseteq\edges$, the set $\Dom(\A)$ satisfies
  the following conditions.
\begin{align}
  \A  &\subseteq \Dom(\A) \tag{D1}\\
  Y_s &\in \Dom(\A) \iff \hat{Y}_s \in \Dom(\A) \tag{D2} \\
  \text{Every}\ e &\in \edges \setminus \Dom(\A)\ \text{is reachable
    from a source}\ \tag{D3} \\
  Y_s &\in \graph(\Dom(\A),s)\ \text{is connected to}\ \hat{Y}_s,
  \forall s\in\sessions \tag{D4}
\end{align}
Furthermore, any set $\B$ satisfying these conditions contains $\Dom(\A)$.
\end{theorem}

Although the authors give this notion of information dominance
in~\cite{HarKle06}, they did  not use it to derive an easily computable
bound. We now formulate a new bound
using information dominance along similar lines to our other bounds and
compare this new bound with ours.

According to~\cite{HarKle06}, the well known linear programming bound
$\R(\Gamma)$ uses a constraint that can be viewed as a restricted
version of information dominance used in the linear programming outer
bound defined in~\cite[Section VIII]{HarKle06}. 

We remark that, for directed acyclic networks, the bound in~\cite[Section VIII]{HarKle06} simply coincides with $\R(\Gamma)$.  It uses $\Gamma$, $\C_1$,
$\C_2$, $\C_3$ and $\C_4$ (as used in $\R(\Gamma)$) together with
information dominance. However, since information dominance is implied
by $\C_1\cap\C_2\cap\C_3\cap\C_4\cap\Gamma$, it does not actually  introduce any new constraints. This can be rigorously justified by
Corollary~\ref{thm:DomSubPsi} proved in this section, since only
polymatroid constraints are used, apart from the constraints
introduced by network demands, to characterize the set $\psi(\cdot)$, and hence $\Dom(\cdot)$ (Theorem
\ref{thm:ID}).

Following our program for developing bounds established in
Sections \ref{sec:mainresults}, we now define maximal
information dominating sets and formulate a bound in terms of
these sets.
\begin{definition}[Maximal Information Dominating Set]\label{def:MaxIDset}
  For a given network $\graph = (\nodes, \edges)$, a set
  $\A\subseteq\edges$ is a maximal information dominating set if
  $\Dom(\A)= \edges$ and no proper subset of $\A$ has the same
  property.
\end{definition}
\begin{lemma}\label{lem:MaxIDset}
  For a given network $\graph = (\nodes, \edges)$ the joint entropy of
  any maximal information dominating set is the same as the joint
  entropy of all source random variables.
\end{lemma}
\begin{IEEEproof}
  First note that the set of all source random variables,
  $Y_{\sessions}$, is a maximal dominating set. Let $U_{\edges}$
  denote set of all edge random variables. Then 
  $$H(U_\edges,
  Y_{\sessions})= H(U_\edges \mid
  Y_\sessions)+H(Y_\sessions)=H(Y_\sessions).$$ 
  Now, let $\B$ be any
  other maximal information dominating set. Then $H(\{U_\edges,
  Y_{\sessions}\}\setminus \mathcal B \mid \mathcal B)=0$ and hence
  $H(U_\edges, Y_{\sessions})= H(\mathcal B)$
\end{IEEEproof}

\begin{theorem}[Information Dominance Bound]\label{thm:IDbound}
  Let $\graph=(\nodes,\edges)$ be a given network with
  network coding constraints. Let $\{U_{\A},Y_{\W^c}\}$ be a
  maximal information dominating set  according to
  Definition~\ref{def:MaxIDset}. Then
  \begin{equation}\label{eq:IDbound}
    \sum_{s \in \W}H(Y_s) \leq  \sum_{e \in \A} c_e.
\end{equation}
\end{theorem}
\begin{IEEEproof}
  The proof is similar to that in  Theorem~\ref{thm:mainresult1},
  by invoking  Lemma~\ref{lem:MaxIDset} and submodularity. 
%  subadditivity of $H\in\Gamma^*$ and
%  $\C_4$.
%where $(a),(b)$ and $(c)$ follows from Lemma \ref{lem:MaxIDset}, subadditivity of $H\in\Gamma^*$ and $\C_4$, respectively.
\end{IEEEproof}

Let $\boldsymbol{\mathcal I}$ be the set of all maximal information dominating sets. Define the \emph{information dominance region} as follows.
\begin{equation}
  \R_{ID} \triangleq \bigcap_{\{U_{\A},Y_{\W^c}\} \in \boldsymbol{\mathcal I}} \left\{(c_e:e
    \in \edges) : \eqref{eq:IDbound}\
    \text{holds}\right\}.
\end{equation}

In the following we establish that $\Dom(\A)\subseteq\psi(\A)$. This will lead us to the conclusion that
$\R_{FD}^{\perp} \subseteq \R_{ID}$. We will proceed by considering each of
the conditions $\mathrm{(D1)}$ -- $\mathrm{(D4)}$ in the definition of
information dominance, and relating them to $\psi$.  
\begin{lemma}\label{lem:dom-psi1}
  Let $\graph^*=(\X,\edges^*)$ be the network FDG of a given network. Then for any $\A \subseteq
  \X$, $\psi(\A)$ satisfies Conditions $\mathrm{(D1)},\mathrm{(D2)}$ of
  $\Dom(\A)$.
\end{lemma}
\begin{IEEEproof}
  By Definition \ref{def:fd1}, the node representing the source
  variable $Y_s$ is in $\psi(\A)$ if and only if any of nodes $\hat
  Y^i_s,i \in b(s)$ representing decoding constraints (i.e., estimated
  source variables) is in $\psi(\A)$. This is equivalent to the
  Condition $\mathrm{(D2)}$ for $\Dom(\A)$. Also note that, by definition, $\A \subseteq \psi(\A)$.
\end{IEEEproof}

\begin{definition}
  For a given FDG $\graph^*=(\X,\edges^*)$ and
  a set of nodes $\psi(\A)\subseteq \X$, the graph
  $\graph^*\setminus \psi(\A)$ contains nodes $\X$
  and edges $\edges^* \setminus \{e:\head{e} \in \psi(\A)\}$.
\end{definition}

Condition $\mathrm{(D3)}$ for $\Dom(\cdot)$ requires every node $A \in \X \setminus \psi(\A)$ to have a directed path from a source node
in $\graph^* \setminus \psi(\A)$.

Note that in~\cite{HarKle06}, it is explicitly assumed that there
exists a path from some source nodes to every edge of a given
network. Without this assumption $\mathrm{(D3)}$ may not be
satisfied. Therefore we impose the same restriction to ensure that
$\psi(\A)$ satisfies $\mathrm{(D3)}$ (this assumption is used in the
proof of Lemma~\ref{lem:dom-psi3} below). It is also assumed
in~\cite{HarKle06} that for every session $s$ there is a path from
node $a(s)$ to $b(s)$ in a given $|\sessions|$-pair communication
network.

\begin{lemma}\label{lem:dom-psi3}
  Let $\graph^*=(\X,\edges^*)$ be a network FDG. Every node in $\X \setminus \psi(\A)$ has a
  directed path from a source node in $\graph^* \setminus \psi(\A)$.
\end{lemma}
\begin{IEEEproof}
  %\comment{AG: I can't understand this proof, since I don't understand the
  %  definition.}
  By assumption (on the network
  model~\cite{HarKle06}), every node in the network FDG
  $\graph^*$ has a directed path from some source node. Now we prove
  that the statement of the lemma is true by contradiction.
  Assume that there exists a node $A \in \X \setminus \psi(\A)$ in $\graph^* \setminus \psi(\A)$
  which has no directed path from any source node. Then it follows
  that every path from any source node to the node $A$ in $\graph^*$
  intersects at least one node from $\psi(\A)$. Then, $A \in \psi(\A)$ and hence there cannot exist
  such a node $A \in \graph^*\setminus \psi(\A)$.
\end{IEEEproof}
\begin{corollary}\label{lem:dom-psi2}
  Let $\graph^*=(\X,\edges^*)$ be a network FDG. Then for any set $\A \in \X$, $\psi(\A)$
  satisfies Condition $\mathrm{(D3)}$ of $\Dom(\A)$.
\end{corollary}

So far we have shown that Conditions $\mathrm{(D1)}-\mathrm{(D3)}$ of
$\Dom(\cdot)$ are satisfied by $\psi(\cdot)$. Now we show that the
Condition $\mathrm{(D4)}$ is equivalent to \fd-separation in
network FDG.
\begin{lemma}\label{lem:dom-psi4}
  For a given network $\graph$, $Y_s$ is connected to $\hat Y_s$ in
  $\graph(\Dom(\A), s)$ if and only if $\A$ does not \fd-separate
  $Y_s$ and $\hat Y_s$ in the network FDG, i.e., Condition $\mathrm{(D4)}$ of $\Dom(\cdot)$ and
  \fd-separation are the same.
\end{lemma}
\begin{IEEEproof}
  By Definition~\ref{def:DomModifiedG}, $\graph(\Dom(\A), s)$ is a
  subgraph of the network obtained by 1) considering the ancestral
  part of $Y_s, \hat Y_s$ and then 2) removing edges in $\A$ and
  subsequently removing all edges which have no path from any
  source. Now, if the edge representing $Y_s$ incoming to the node
  $a(s)$ is connected to an edge representing $\hat Y_s$ outgoing from
  any node in $b(s)$ in $\graph(\Dom(\A), s)$ then in the
   network FDG, $\A$ does not \fd-separate $Y_s$
  and $\hat Y_s$.  Also, in network FDG, if $\A$ does not
  \fd-separate $Y_s$ and $\hat Y_s$ then there exists a connection
  between $Y_s$ and $\hat Y_s$ in $\graph(\Dom(\A), s)$.
\end{IEEEproof}

This leads us to the following conclusions. By Lemmas~\ref{lem:dom-psi1} and Corollary~\ref{lem:dom-psi2}, Conditions
  $\mathrm{(D1)}$ -- $\mathrm{(D3)}$ are satisfied by our notion of
  functional dependence in Definition~\ref{def:fd1}. By
  Lemma~\ref{lem:dom-psi4}, Condition $\mathrm{(D4)}$ is equivalent to
  \fd-separation, which is employed in Definition~\ref{def:fd1} and hence

\begin{corollary}\label{thm:DomSubPsi}
  \begin{equation}
    \psi(\A) \subseteq \Dom(\A).
  \end{equation}
\end{corollary}
%\begin{IEEEproof}
%\end{IEEEproof}
\begin{corollary}\label{thm:ID is DI}
  \begin{equation}
    \R_{FD}^{\perp} \subseteq \R_{ID}.
  \end{equation}
\end{corollary}

%\begin{IEEEproof}
  The corollary follows from Corollary~\ref{thm:DomSubPsi} and the fact
  that the information dominance bound (Theorem~\ref{thm:IDbound}) and
  the functional dependence bound for independent sources
  (Theorem~\ref{thm:mainresult2}), apart from characterization of
  $\Dom(\cdot)$ and $\psi(\cdot)$, use the same arguments.
%\end{IEEEproof}

\subsection{Progressive $d$-Separating Edge-Set Bound}
In~\cite{KraSav06} the authors describe a procedure to determine
whether a given set of edges bounds the capacity of the given
network. The progressive $d$-separating edge-set (\pde)
bound uses the concept of \fd-separation~\cite{Kra98}. The
results are given for general cyclic multi-source multi-sink networks
with noisy channels.
{
\begin{definition}[\pde\  Procedure]\label{def:PdE}
  The \pde\  procedure determines whether a given set of edges
  $\A$ bounds the capacity of information flow for sources
  $Y_{\W} \subseteq Y_{\sessions}$ for some ordering of the
  elements of $\W$ defined by the relation $\prec$
  as follows.
  \begin{enumerate}
%% What does it mean to "move backward" is this directed or undirected?
  \item In the functional dependence graph\footnote{The definition of
      a functional dependence graph used here is different from
      that defined in Section \ref{sec:mainresults}, see~\cite{KraSav06}.} of the
    given network, remove all vertices and edges in $\graph$
    except those encountered when moving backward one or more edges
    starting from any of the vertices representing $\A,
    \{Y_{s_{i(\prec)}}: s_i \in \W\}$ and $\{\hat Y_{s_{i(\prec)}}: s_i
    \in \W\}$. Further remove edges coming out of vertices
    representing $\A$ and $Y_{\W^c}$ and successively remove edges
    coming out of vertices and on cycles that have no incoming edges,
    excepting source vertices. Set $i=1$.
  \item (Iterations) If $Y_{s_{i(\prec)}}$ is not disconnected (in an
    undirected sense) from all of its estimates $\hat
    Y_{s_{i(\prec)}}$, then STOP (one has no bound). Else if
    $Y_{s_{i(\prec)}}$ is disconnected (in an undirected sense) from
    one of its estimates then: (a) remove the edges coming out of the
    vertex representing $Y_{s_{i(\prec)}}$. (b) Successively remove
    edges coming out of vertices and edges coming out of vertices that
    have no paths from source vertices.

  \item (Termination and Bound) Increment $i$. If $i \leq |\W|$ go to
    Step 2. If $i=|\W|+1$
    \begin{equation}
      \sum_{s \in \W} H(Y_s) \leq \sum_{e \in \A(\W,\prec)} c_{e}.
    \end{equation}
    where $\A(\W,\prec)$ is referred as a \pde\ set.
\end{enumerate}
\end{definition}

\begin{theorem}\label{thm:PdE bound}
The progressive $d$-separating edge-set bound is
\begin{equation}\label{eq:pdebound}
\sum_{s \in \W} H(Y_s) \leq \sum_{e \in \A(\W,\prec)} c_{e}
\end{equation}
where $\boldsymbol{\A}(\W,\prec)$ is the collection of subsets of
$\edges$ that are \pde\ sets (Definition \ref{def:PdE}) for $\W$ under
the ordering relation $\prec$.
\end{theorem}
The \emph{progressive d-separating edge-set region} is
\begin{equation}
\R_{PdE} \triangleq \bigcap_{\W \subseteq \sessions, \prec} \left\{(c_e:e
\in \edges): \eqref{eq:pdebound}\ \text{holds}\right\}.
\end{equation}
}

From the definitions of the network sharing bound and the P\emph{d}E bound it can be noted that both bounds depend
on a choice of source ordering and to compute the tightest bounds all
possible orderings have to be considered. Also note that determination
of the tightest \pde\ sets involves exhaustively searching over all
subsets of edges for a given source ordering. In contrast, we will use
structural properties of functional dependence to efficiently compute
all network bottlenecks, namely the maximal irreducible sets.

\begin{theorem}\label{thm:IDsubPdE}
\begin{equation}
  \R_{FD}^{\perp}\subseteq \R_{PdE}.
\end{equation}
Furthermore, there exists a network such that the inclusion is strict.
\end{theorem}
\begin{IEEEproof}
  Let $\A$ be a progressive $d$-separating edge-set bounding the rate
  with respect to $Y_{\W}$ for a given network. Then we prove that
  $\psi(\A,Y_{\W^c})=\X$ in network FDG. The rest follows from
  Theorem~\ref{thm:mainresult2}.

  First note that the Step~1(a) in Definition \ref{def:PdE} considers
  ancestral part of $\{\A,Y_{\W},\hat{Y}_{\W}\}$ and Step~1(b) removes edges outgoing
  from nodes in $\A,Y_{\W^c}$ and subsequently removes nodes and edges
  with no incoming edges and nodes respectively (except for source
  nodes). Denote the resulting graph by $\graph'$. Step~2 checks
  connectivity of $Y_s:s \in \W$ and $\hat Y_s:s \in \W$ in $\graph'$
  in iterative manner with respect to some $\prec$.

  In contrast, Definition~\ref{def:fd1} first removes edges outgoing
  from nodes in $\A, Y_{\W^c}$ and successively removes nodes and
  edges with no incoming edges and nodes respectively. In the second
  stage, it checks connectivity of each $Y_s,s \in \W$ with $\hat Y_s,
  S \in \W$ in $\graph^*_{\An(\A, Y_{\W^c})}$ in iterative manner. But
  note that
  \begin{equation*}
    \graph^*_{\An(\A, Y_{\W^c},Y_s,\hat Y^{i}_s: i \in b(s))} \subseteq \graph', s \in \W.
  \end{equation*}
  Hence, if $Y_s:s \in \W$ and $\hat Y_s:s \in \W$ are disconnected in
  $\graph'$ then they are disconnected in $\graph_{\An(\A, Y_{\W^c})}$.
  Thus if a progressive $d$-separating edge-set $\A$ bounds the rate
  of the sources $Y_{\W}$ for a given network then $Y_{\W} \subseteq
  \psi(\A)$ which implies
  $\psi(\A,Y_{\W^c})=\psi(\A,Y_{\W},Y_{\W^c})=\X$ (since
  $\psi(Y_{\sessions})=\X$) and hence $\psi(\A,Y_{\W^c})=\X$
  in  network FDG. Strict inclusion is demonstrated in the following example.
\end{IEEEproof}

{
\begin{example}\label{ex:nspde}
  Figure \ref{fig:exp} shows a three-layer network. Note that source pairs $Y_1,Y_2$ and $Y_3,Y_4$ form two
  butterfly networks. We show that there exists a maximal irreducible set which is strictly smaller than a P\emph{d}E set for bounding the sum-rate capacity of all sources.
  \begin{figure}[htbp]
    \centering
    \includegraphics[scale=0.9]{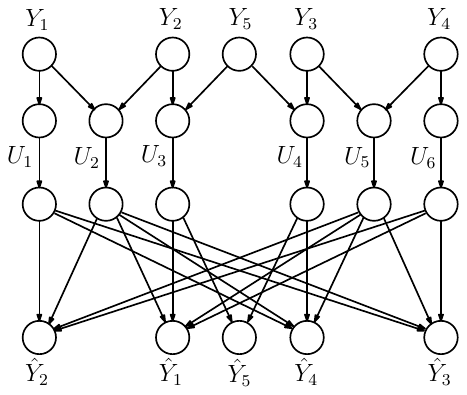}
    \caption{A network example.}\label{fig:exp}
  \end{figure}

  The sum-rate bound, Theorem \ref{thm:mainresult2}, for any proper subset of
  the sources is identical to P\emph{d}E bound, however, the set $\{U_2,U_3,U_4,U_5\}$ is a maximal irreducible set
  yielding
  \begin{equation*}%\label{eq:pdeExp3}
    \sum_{s=1}^5 H(Y_s) \leq c_3+c_3+c_4+c_5.
  \end{equation*}

 Note that, for $s \in \{1,2,3,4,5\}$,  $Y_s$ and $\hat Y_s$ are
  \fd-separated by $\{U_2,U_3,U_4,U_5\}$ in the subgraph
  $\bar{\graph^*}_{\An(Y_s, \hat Y_s, U_2,U_3,U_4,U_5)}$ of
  network FDG. One can also check from Figure~\ref{fig:exp1} that removing
  $\{U_2,U_3,U_4,U_5\}$ for P\emph{d}E
  bound does not disconnect the sources $Y_1, Y_2, Y_3, Y_4$ from their respective sinks.
  Also note that all source variables are in $\Dom(2,3,4,5)$ and hence
  the information dominance bound is also tighter than the P\emph{d}E
  bound for the network in Figure~\ref{fig:exp}.
    \begin{figure}[htbp]
    \centering
    \includegraphics[scale=0.9]{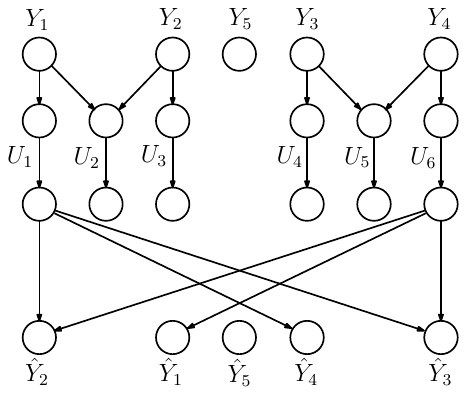}
    \caption{Removing outgoing edges of $\{U_2,U_3,U_4,U_5\}$ in the
      network of Figure \ref{fig:exp}.}\label{fig:exp1}
  \end{figure}
\end{example}
}

Close inspection of Definition~\ref{def:PdE} reveals that
\fd-separation is weaker in the \pde\ bound since it does not consider
the ancestral part of $\A,\B,\C$ when using the \fd-separation
criteria to check $\A \bot \B |\C$. 
The \pde\ bound can be therefore strengthened by modifying it to
consider the ancestral part of $\{Y_k,\hat Y_k, \A\}$. The resulting
improved \pde\ bound would be the same as our
bound for independent sources, Theorem \ref{thm:mainresult2}.

%%%%%%%%%%%%%%%%%%%%%%%%%%%%%%%%%%
\section{Conclusion} \label{sec: conclusion} 
Explicit characterization and computation of the multi-source network coding capacity region
requires determination of the set of all entropic vectors
$\Gamma^{*}$, which is known to be an extremely hard problem. The best
known outer bound can in principle be computed using a linear
programming approach. In practice this is infeasible due to an
exponential growth in the number of constraints and variables with the
network size. We extended previous notions of functional dependence
graphs to accommodate not only cyclic graphs, but more abstract
notions of independence. In particular we considered polymatroidal
functions, and demonstrated efficient and systematic
methods to find functional dependencies implied by the given local
dependencies.  This led to one of our main results, which was a new,
easily computable outer bound, based on characterization of all
implied functional dependencies. %We also gave new outer bounds
%$\R_{\text{cs}} (\bar\Gamma^*)$ and $\R_{\text{cs}}(\Gamma)$ for
%networks with correlated source and 
We showed that the easily
computable functional dependence bound is indeed an outer bound on the
capacity region of general multicast networks with correlated sources. We
extended the notion of irreducible sets for networks with independent
sources and formulated a tighter outer bound for such networks. We
compared the tightness of our proposed bounds with other existing
bounds. We showed that our proposed bounds improve on the cut-set
bound, the network sharing bound, a new bound derived from information
dominance, and the \pde\ bound. Finally, we showed how to make a minor
modification of the \pde\ bound, tightening it to coincide with our bound.

\appendix[Maximal irreducible sets for acyclic Graphs]\label{sec:Rcs}
In a directed acyclic graph, let $\An(\A)$ denote the set of ancestral nodes,
i.e., for every node $a\in\An(\A)$, %$a \not \in \A$ and 
there is a directed path from $a$
to some $b\in\A$. Of particular interest are the maximal irreducible sets:
\begin{definition}\label{def:MaxIrrSetA}
  An irreducible set $\A$ is \emph{maximal} in an acyclic FDG
  $\graph^*=(\X,\edges^*)$ if
  $\X\setminus\varphi(\A)\setminus\An(\A) \defined
  (\X\setminus\varphi(\A))\setminus\An(\A) =\emptyset$, and no
  proper subset of $\A$ has the same property.
\end{definition}

Note that for acyclic graphs, every subset of a maximal irreducible
set is irreducible. %Conversely, every irreducible set is a subset of
%some maximal irreducible set \cite{GriGra08}. 
Irreducible sets can be
augmented in the following way.
\begin{corollary}[Augmentation]\label{lem:augment}
  Let $\A\subseteq\nodes$ in an acyclic FDG $\graph^*=(\X,\edges^*)$. Let
  $\B=\X\setminus\varphi(\A)\setminus\An(\A)$.  Then $\A\cup\{b\}$ is
  irreducible for every $b\in \B$.
\end{corollary}

This suggests a process of recursive augmentation to find all maximal
irreducible sets in an acyclic FDG (a similar process of augmentation
was used in \cite{GriGra08}).  Let $\graph^*$ be a topologically
sorted\footnote{Here, we assume that  if there is a directed edge
  from node $i$ to $j$, then $i\prec j$ \cite[Proposition
  11.5]{Yeu02}.} acyclic FDG
$\graph^*=(\{0,1,2,\dots\},\edges^*)$. Its maximal irreducible sets can be
found recursively via $\textbf{AllMaxSetsA}(\graph^*,\{\})$ in Algorithm
\ref{alg:augment}. In fact, $\textbf{AllMaxSetsA}(\graph^*,\A)$
finds all maximal irreducible sets containing $\A$ given that the set $\A$ is an irreducible set and $\graph^*$ is finite.
\begin{algorithm}[htbp]
\caption{\textbf{AllMaxSetsA}($\graph,\A$)}\label{alg:augment}
  \begin{algorithmic}[1]
    \REQUIRE $\graph^*=(\X,\edges^*), \A\subseteq\X$
    \STATE
    $\B\leftarrow\X\setminus\varphi(\A)\setminus\An(\A)$ \label{stp:A1-1}
    \IF{$\B\neq\emptyset$}
    \STATE Output $\{\textbf{AllMaxSetsA}(\graph^*,\A\cup \{b\} ): b\in\B \}$ \label{stp:A1-3}
    \ELSE
    \STATE Output $\A$
    \ENDIF
  \end{algorithmic}
\end{algorithm}

The actual number of calls of the function \textbf{AllMaxSetsA}($\cdot,\cdot$) to compute all maximal irreducible sets depends on the topology of the FDG. For example, for a line FDG $\mathcal G^* =(\X,\edges^*)$ with $\X =\{i: 1\leq i \leq n\}$ and $\edges^* =\{(i,i+1): 1\leq i < n\}$, the number of times the function \textbf{AllMaxSetsA}($\cdot,\cdot$) called is only $n+1$ (linear in the order of $\mathcal G^*$).

For an acyclic FDG $\graph^*$, let $\sessions$ denote the set of nodes
which do not have any parent nodes. Clearly, $\sessions$ is a maximal
irreducible set.
 %Also note that, %$\varphi(\sessions)= \nodes$ and hence $h(\sessions)=h(\nodes)\geq h(A)$.
%\begin{corollary}\label{lem:gequal}
  Let $\sessions$ be the set of nodes without a parent node in a given
  acyclic FDG $\graph^*=(\X,\edges^*)$ and let $\A$ be another maximal
  irreducible set then
%\begin{equation}
$h(\sessions) \geq h(\A)$ since $\varphi(\sessions)= \X$ and hence $h(\sessions)=h(\X)\geq h(\A)$.

\section*{Acknowledgement}
This work was supported in part by the Australian Government
under ARC grant DP150103658.

%%%%%%%%%%%%%%%%%%%%%%%%%%%%%%%%%%%%%%

\bibliographystyle{ieeetr}
\bibliography{network}

\end{document}